\documentclass[aps,prd,twocolumn,showpacs,nofootinbib]{revtex4-1}

\usepackage{graphicx}
\usepackage{amsmath}
\usepackage{bm}
\usepackage{slashed}
\usepackage{epsfig}
\usepackage{amsfonts}
\usepackage{epstopdf}
\usepackage{color}
\usepackage{extarrows}
\usepackage{multirow}
\usepackage[utf8]{inputenc}
\usepackage{hyperref}

\allowdisplaybreaks

\begin{document}

\title{Next-to-leading order QCD corrections to $B_c^*\to J/\psi$ form factors}

\author{Qin Chang$^{a}$}
\author{Wei Tao$^{a}$}
\email{taowei@htu.edu.cn}
\author{Zhen-Jun Xiao$^{b}$}
\author{Ruilin Zhu$^{b}$}

\affiliation{$^a$ Institute of Particle and Nuclear Physics,
	Henan Normal University, Xinxiang 453007, China\\
	$^b$ Department of Physics and Institute of Theoretical Physics, Nanjing Normal University, Nanjing, Jiangsu 210023, China}

\date{\today}

\begin{abstract}
	
Within the framework of Non-Relativistic Quantum Chromodynamics (NRQCD) factorization, we calculate the next-to-leading order (NLO) perturbative QCD corrections to the form factors for  the  semileptonic decays of  $B_c^*$  into $J/\psi$  via  (axial-)vector and (axial-)tensor currents.
We obtain the complete analytical results for the form factors up to NLO, and provide their asymptotic expressions in the hierarchical heavy quark limit. 
The NLO corrections are found to be both significant and convergent in the relatively small squared transfer momentum ($q^2$) region, while also reducing the dependence on the renormalization scale $\mu$.
Finally, the theoretical predictions for $B_c^*\to J/\psi$ form factors over the full $q^2$ range are provided.

\end{abstract}

\maketitle

\section{Introduction}

In the Standard Model of particle physics, the bottom-charm meson is the only meson  composed of two different heavy flavor quarks, and therefore plays an important role in the study of strong  and electroweak interactions.
The ground-state pseudoscalar $B_c$ meson was first experimentally discovered through the semileptonic decay $B_c \to J/\psi+l {\nu}_l$~\cite{CDF:1998ihx}. However, the ground-state vector meson $ B_c^* $ has not yet been observed experimentally~\cite{CMS:2019uhm,LHCb:2019bem}. Therefore, studying the form factors for the semileptonic decay of $ B_c^* $ into $J/\psi $  will provide important reference information for the future experimental discovery of  $ B_c^* $.

Currently, only the light-front quark model (LFQM) has systematically calculated the vector and axial-vector form factors for $B_c^* \to J/\psi$~\cite{Chang:2018mva,Chang:2018sud,Wang:2024cyi,Chang:2019obq,Chang:2020xvu}. Although the HPQCD collaboration has performed lattice QCD calculations of the vector and axial-vector  form factors for the semileptonic decay of the pseudoscalar meson $B_c$ into $J/\psi (\eta_c)$~\cite{Colquhoun:2016osw,Harrison:2020gvo}, no lattice results are available for the semileptonic decay of the vector meson $ B_c^*$.

NRQCD effective theory factorizes physical quantities into short-distance coefficients multiplied by long-distance matrix elements and performs a double expansion in both the strong coupling constant $\alpha_s$ and the relative velocity $v$ between heavy quarks in a meson, which makes it a powerful tool for systematically studying the decay and production of bottom-charm mesons~\cite{Bodwin:1994jh}. 
Using NRQCD, G. Bell et al. performed the first one-loop calculation of the vector and tensor form factors for  $B_c\to \eta_c$ in 2007~\cite{Bell:2006tz}. Since 2011, one-loop calculations of the vector, axial-vector, tensor, and axial-tensor form factors for $B_c \to J/\psi (\eta_c)$ have been successively completed~\cite{Qiao:2011yz,Qiao:2012vt,Tao:2022yur}. From 2017 onward, relativistic corrections to various form factors for the semileptonic decay of $B_c$ into S-wave or P-wave charmonium have also been computed~\cite{Zhu:2017lqu,Zhu:2017lwi,Shen:2021dat}. However, for $B_c^* \to J/\psi$, only tree-level results for the vector and axial-vector form factors are currently available~\cite{Geng:2023ffc}.

To obtain more precise and reliable theoretical predictions, this paper aims to further investigate the one-loop corrections to the vector, axial-vector, tensor, and axial-tensor  form factors for $B_c^* \to J/\psi+l {\nu}_l$ using the NRQCD effective theory.
This calculation not only helps to test the convergence of the perturbative series expansion and the dependence on the renormalization scale, but also aids in testing the Standard Model with higher precision and in determining its fundamental parameters. 
Moreover, calculating higher-order QCD corrections to the (axial-)tensor form factors will also contribute to the exploration of new physics beyond the Standard Model.

The remainder of the paper is organized as follows. In Sec.~\ref{DEFINITION}, we define the (axial-)vector and (axial-)tensor form factors for $B_c^*\to J/\psi$. In Sec.~\ref{CALCULATIONPROCESS}, we describe the calculation procedure for NLO corrections to form factors within NRQCD. In Sec.~\ref{ANALYTICALRESULTS} and the appendix~\ref{appendix123}, we provide the analytical results for the LO and NLO form factors, respectively. In Sec.~\ref{NUMERICALRESULTS}, we present the numerical results up to NLO and the theoretical predictions by  NRQCD + Lattice + Z-series for $B_c^*\to J/\psi$ form factors. We summarize in Sec.~\ref{SUMMARY}.

\section{Definition}\label{DEFINITION}

In QCD, the vector and axial-vector current form factors of $B_c^*$ decay into $J/\psi$ can be defined as~\cite{Chang:2019obq,Chang:2020xvu}
\begin{align}
&\left\langle J/\psi\left(\epsilon^{\prime }, p^{\prime }\right)\left|\bar{b} \gamma_\mu c\right| B_c^*\left(\epsilon, p\right)\right\rangle\nonumber\\
= & -\left(\epsilon \cdot \epsilon^{ \prime *}\right)\left[P_\mu V_1\left(q^2\right)-q_\mu V_2\left(q^2\right)\right] -\left(\epsilon\cdot q\right) \epsilon_\mu^{ \prime *} V_3\left(q^2\right)\nonumber\\&+\left(\epsilon^{\prime *} \cdot q\right) \epsilon_\mu V_4\left(q^2\right)+\left(\epsilon \cdot q\right)\left(\epsilon^{ \prime *} \cdot q\right)\left[\left(\frac{P_\mu }{M^{ 2}-M^{\prime 2}}\right.\right.\nonumber\\&\left.\left.-\frac{q_\mu}{q^2} \right) V_5\left(q^2\right)+\frac{q_\mu}{q^2}  V_6\left(q^2\right)\right],
\end{align}
and
\begin{align}
&\left\langle J/\psi\left(\epsilon^{\prime }, p^{\prime }\right)\left|\bar{b} \gamma_\mu \gamma_5 c\right| B_c^*\left(\epsilon, p\right)\right\rangle\nonumber\\= 
&
i \varepsilon_{\mu \nu \alpha \beta} \epsilon^{\alpha} \epsilon^{ \prime * \beta}\left[P^\nu A_1\left(q^2\right)- q^\nu A_2\left(q^2\right)\right] \nonumber\\
& +\frac{i \varepsilon_{\mu \nu \alpha \beta} P^\alpha q^\beta}{M^{ 2}-M^{ \prime 2}}\left[\epsilon^{\prime *} \cdot q \epsilon^{ \nu} A_3\left(q^2\right)-\epsilon \cdot q \epsilon^{ \prime * \nu} A_4\left(q^2\right)\right]
\nonumber\\
&
- \frac{i \varepsilon_{\rho \nu \alpha \beta}\epsilon^{\alpha} \epsilon^{ \prime * \beta}P^\nu q^{ \rho}}{M^{ 2}-M^{ \prime 2}}\left[P_{ \mu} A_5\left(q^2\right)-q_{ \mu} A_6\left(q^2\right)\right]
.
\end{align}

Similarly, the tensor and axial-tensor current form factors of $B_c^*$ decay into $J/\psi$ can be defined as
\begin{align}\label{tdef}
&\left\langle J/\psi\left(\epsilon^{\prime }, p^{\prime }\right)\left|\bar{b} \sigma_{\mu\nu} q^\nu c\right| B_c^*\left(\epsilon, p\right)\right\rangle\nonumber\\
= 
& -i\left(\epsilon \cdot \epsilon^{ \prime *}\right)\left[P_\mu T_1\left(q^2\right)-q_\mu T_2\left(q^2\right)\right](M+M^{\prime})\nonumber\\& -i\left[\left(\epsilon\cdot q\right) \epsilon_\mu^{ \prime *} T_3\left(q^2\right)-\left(\epsilon^{\prime *} \cdot q\right) \epsilon_\mu T_4\left(q^2\right)\right](M+M^{\prime})\nonumber\\&+i\frac{\left(\epsilon \cdot q\right)\left(\epsilon^{ \prime *} \cdot q\right)}{M+M^{\prime }}\left[{P_\mu } T_5\left(q^2\right)+{q_\mu}  T_6\left(q^2\right)\right],
\end{align}
and
\begin{align}\label{t5def}
&\left\langle J/\psi\left(\epsilon^{\prime }, p^{\prime }\right)\left|\bar{b}  \sigma_{\mu\nu}\gamma_5 q^\nu  c\right| B_c^*\left(\epsilon, p\right)\right\rangle\nonumber\\= &
- \varepsilon_{\mu \nu \alpha \beta} \epsilon^{\alpha} \epsilon^{ \prime * \beta}\left[P^\nu T'_1\left(q^2\right)+ q^\nu T'_2\left(q^2\right)\right](M+M^{ \prime }) \nonumber\\
& +\frac{ \varepsilon_{\mu \nu \alpha \beta} P^\alpha q^\beta}{M+M^{ \prime }}\left[\epsilon^{\prime *} \cdot q \epsilon^{ \nu} T'_3\left(q^2\right)+\epsilon \cdot q \epsilon^{ \prime * \nu} T'_4\left(q^2\right)\right]
\nonumber\\&
-\frac{\varepsilon_{\rho \nu \alpha \beta}\epsilon^{\alpha} \epsilon^{ \prime * \beta}P^\nu q^{ \rho}}{M+M^{ \prime }}\left[P_{ \mu}T'_5\left(q^2\right)-q_{ \mu}T'_6\left(q^2\right)\right]
.
\end{align}
Here, $\sigma_{\mu\nu}=\frac{i}{2}\left(\gamma_\mu\gamma_\nu-\gamma_\nu\gamma_\mu\right)$,   $P=p+p'$,  $M(M')$ is the  mass of $B_c^*(J/\psi)$. And  $q=p-p'$ is the transfer momentum, which  satisfies the physical constraint: $0\leq q^2\leq (M-M')^2$.

Note, for tensor and axial-tensor cases, 
multiplying both sides of Eqs.~\eqref{tdef} and \eqref{t5def} by $q^\mu$, one can obtain the following identities:
\begin{align}
&0\equiv(M^2-M'^2)T_1-q^2 T_2,\\
&0\equiv(M+M')^2(T_3-T_4)-(M^2-M'^2)T_5-q^2 T_6,\\
&0\equiv(M+M')^2 T'_1+(M^2-M'^2)T'_5-q^2 T'_6.
\end{align}
By separately computing each form factor, we have verified the above relations. Moreover, the actual calculation results show that $A_5\equiv A_6\equiv T'_1\equiv T'_5\equiv T'_6\equiv 0$. Therefore, we only need to present the results for the form factors $V_{1,2,3,4,5,6}$, $A_{1,2,3,4}$, $T_{2,3,4,6}$ and $T'_{2,3,4}$ in this paper.

\section{Calculation process}\label{CALCULATIONPROCESS}

Within the NRQCD factorization framework, 
the form factors (semileptonic decay amplitudes)  in QCD can be factorized  as the product of short-distance coefficients and NRQCD long-distance matrix elements (LDMEs). The NRQCD LDMEs encapsulate the contributions from soft, ultrasoft, and potential momentum regions of the QCD amplitudes and are generally regarded as universal nonperturbative quantities.  These LDMEs are expanded in terms of the relative velocity between the quarks inside the mesons, and at the leading order of the velocity expansion, they reduce to the mesons’ wave functions at the origin~\cite{Qiao:2011yz,Tao:2022yur}. In contrast, the short-distance coefficients correspond to the contributions from the hard momentum region of the QCD amplitudes and can be calculated up to higher orders in the strong coupling constant. Therefore, the main task of this work is the one-loop perturbative calculation of the short-distance coefficients (hard QCD amplitudes).

\begin{figure}[htbp]
	\centering
	\includegraphics[width=0.45\textwidth]{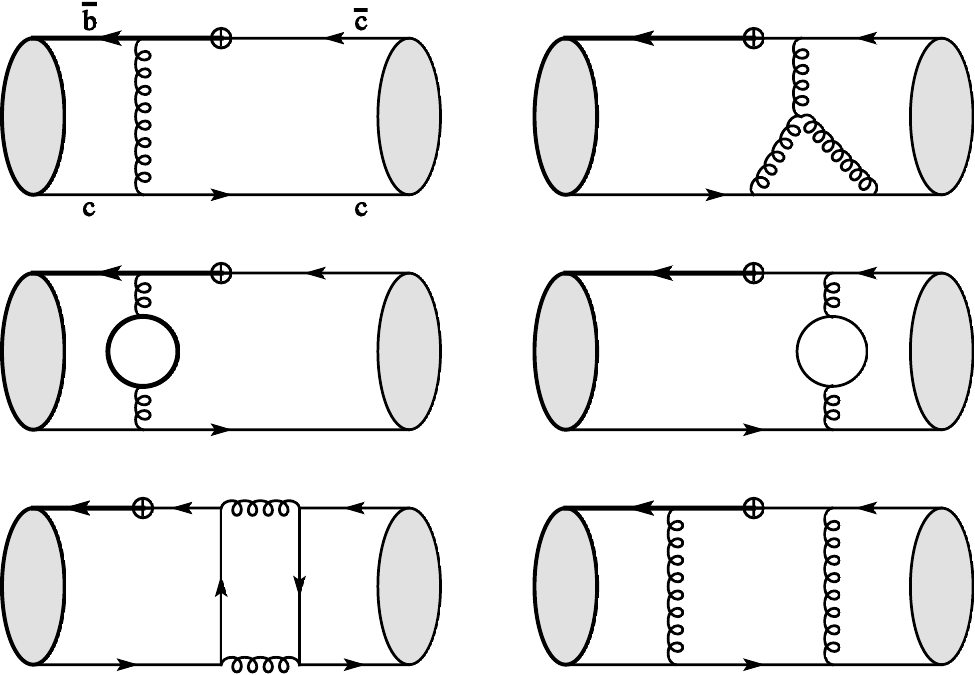}
	\caption{Representative Feynman diagrams at tree and one-loop levels, where ``$\oplus$''  denotes a certain heavy flavor-changing current vertex and the thick (thin) solid closed circle represents the bottom (charm) quark loop. The diagram on the left of the last row is a flavor-singlet diagram.}
	\label{fig:bc2cctree1loop}
\end{figure}

We use \texttt{FeynCalc}~\cite{Shtabovenko:2023idz} to generate Feynman diagrams for the process $c\bar b\to c \bar c W^*$, and consider $n_l$ massless quarks, $n_b$ bottom quarks with mass $m_b$ and $n_c$ charm quarks with mass $m_c$ possibly appearing in the quark loop. At the tree level, there are 2 diagrams, while at the one-loop order, there are 37 diagrams, of which 6 flavor-singlet diagrams contribute a total of 0 to $B_c^*\to J/\psi$ form factors~\cite{Bell:2006tz,Tao:2022yur}.
Some representative Feynman diagrams are shown in Fig.~\ref{fig:bc2cctree1loop}.

Then we employ the perturbative matching procedure and the covariant projection technique~\cite{Petrelli:1997ge,Bell:2006tz,Jia:2024ini} to extract the hard QCD amplitudes for form factors.
With the aid of \texttt{FeynCalc}, the form factors can be expressed in terms of scalar products of momenta after performing Dirac matrix simplification, index contraction, color algebra simplification, and trace calculation. 
For trace computation of the fermion chain containing one $\gamma_5$, we adopt the fixed reading point $\gamma_5$ scheme~\cite{Shtabovenko:2023idz,Korner:1991sx,Larin:1993tq,Moch:2015usa}, where the reading point is determined as follows.
If the fermion chain contains the axial-vector or axial-tensor current vertex,  the following replacement rule is applied to it:
\begin{align}
\text{Trace}\left(a\cdot\Gamma\cdot b\right)\to \text{Trace}\left(\frac{\Gamma\cdot b\cdot a+b\cdot a\cdot\Gamma}{2}\right),
\end{align}
where  $\Gamma=\gamma_\mu\gamma_5$ or $\Gamma=\sigma_{\mu\nu}\gamma_5$.
Otherwise,  the following replacement rule is applied to the fermion chain:
\begin{align}
\text{Trace}\left(a\cdot\gamma_5\cdot b\right)\to \text{Trace}\left(b\cdot a\cdot\gamma_5\right).
\end{align}
For the fermion chain containing an even number of $\gamma_5$, the naive $\gamma_5$ scheme is used~\cite{Shtabovenko:2023idz}.

By invoking the \texttt{FeynCalc}'s function \texttt{TID}, the form factors from  one-loop diagrams are ultimately expressed in terms of the standard Passarino-Veltman functions: $A_0,B_0,C_0,C_1,D_0,E_0$, where the 5-point function $E_0$ can be further reduced to $A_0,B_0,C_0$ with the aid of IBP reduction~\cite{Chetyrkin:1981qh}.
Then the remaining functions $A_0,B_0,C_0,C_1,D_0$ can be analytically calculated by utilizing \texttt{Package-X}~\cite{Patel:2016fam}.
In order to obtain the asymptotic results of the form factors in the hierarchical heavy quark limit (by expanding in small $m_c$ and taking the leading-order terms), we use the \texttt{Mathematica} function \texttt{Series}  combined with some in-house codes to perform the power expansion in $m_c$ for the complicated \texttt{PolyLog} functions appearing in the analytical expressions of $A_0,B_0,C_0,C_1,D_0$.

In dimensional regularization, the one-loop amplitudes contributing to form factors contain  
ultra-violet (UV) and infra-red (IR) divergence poles.
To cancel these divergences and obtain the finite results for the form factors, we need to perform the renormalization procedure, which requires summing contributions from the one-loop diagrams and the tree diagrams inserted with one $\alpha_s$-order counterterm vertex. 
The counterterm diagrams involve the QCD coupling renormalization constant in the modified-minimal-subtraction (${\overline{\mathrm{MS}}}$) scheme ~\cite{Mitov:2006xs,Chetyrkin:1997un,vanRitbergen:1997va}, the QCD heavy quark field (mass)  renormalization constant in the on-shell (OS) scheme~\cite{Fael:2020bgs}, and the QCD heavy flavor-changing current OS renormalization constant $Z_{J}^\mathrm{{OS}}$.
For vector and axial-vector currents, $Z_{J}^\mathrm{{OS}}\equiv 1$.
For tensor and axial-tensor currents, the expression of $Z_{J}^\mathrm{{OS}}$ can be obtained from our previous research~\cite{Tao:2023vvf,Tao:2023pzv}, which reads~\footnote{Note that in our earlier work~\cite{Tao:2022yur}, the QCD heavy flavor-changing (axial-)tensor current ${\overline{\mathrm{MS}}}$ renormalization constant $Z_{J}^{\overline{\mathrm{MS}}}$ was adopted, which differs from the on-shell renormalization constant  $Z_{J}^\mathrm{{OS}}$ used in this paper.}:
\begin{align}
Z_J^\mathrm{OS}&=1+\frac{\alpha_s C_F}{4\pi}\left(\frac{1}{\epsilon}-\frac{2x \ln x}{1+x}+2\ln y+{\cal O}(\epsilon)\right)\nonumber\\
&~~~~+{\cal O}(\alpha_s^2).
\end{align}
Here, the following dimensionless variables have been defined and will be used throughout the paper:
\begin{align}
x=\frac{m_c}{m_b},~~~~~
y=\frac{\mu}{m_b},~~~~~
s=\frac{1}{1-\frac{q^2}{m_b^2}}.
\end{align}
It is worth mentioning that since the anomalous dimensions of the NRQCD heavy flavor-changing currents arise starting from ${\cal O}(\alpha_s^2)$~\cite{Tao:2022qxa,Tao:2023pzv}, the NLO calculation of  the form factors does not involve the NRQCD heavy flavor-changing current renormalization.

To verify the correctness of our calculations, the following checks have been performed.  After summing all contributions, the NLO results of the form factors are found to be finite and satisfy renormalization group invariance~\cite{Tao:2023mtw,Tao:2023vvf,Tao:2023pzv}. Furthermore, all calculations are carried out with a general gauge parameter, and the NLO results of the form factors are confirmed to be independent of it.  Additionally, the analytical results have been cross-checked against the numerical multi-loop method---expressing the form factors in terms of Feynman integrals, reducing them to master integrals, and computing the master integrals using the \texttt{AMFlow} package~\cite{Liu:2017jxz,Liu:2022chg}---yielding perfect agreement.

\section{Analytical results}\label{ANALYTICALRESULTS}
As follows, we present the analytical calculation results for the $B_c^*\to J/\psi$ form factors, involving the vector, axial-vector, tensor, and axial-tensor currents, up to NLO of $\alpha_s$ within the NRQCD factorization framework.
  
At the leading order (LO), the form factors can be expressed as
\begin{align}\label{completeLO}
&V_1=\frac{16\sqrt{2} \pi\alpha _s C_F s^2(1+x)^{\frac{5}{2}}\Psi_{B_c^*}(0)\Psi_{J/\psi}(0)}{m_b^3 x^{\frac{3}{2}}(1+s(x-2)x)^2},\nonumber\\
&V_2=A_2=T_2=\frac{1-x}{1+x}V_1=\frac{2(1+x)}{1+3x}T_6,\nonumber\\
&T_3=\frac{1+x}{2x}T_4=\frac{-1+s(4+10 x+3 x^2)}{2s(1+x)(1+3x)}V_1,\nonumber\\
&V_1=\frac{V_3}{2}=A_1=\frac{1+x}{4x}V_4=\frac{s(1+x) (1+3x) }{1+4s x+3s x^2}T'_2,\nonumber\\
&T'_4=\frac{1+x}{2}T'_3=\frac{1+3x}{2(1+x)}V_1,\nonumber\\
&V_5=V_6=A_3=A_4=0,
\end{align}
where $\Psi_{B_c^*}(0)$ and $\Psi_{J/\psi}(0)$ are the wave functions at the origin for the mesons $B_c^*$ and $J/\psi$, respectively. 
In the hierarchical heavy quark limit, the form factors in Eq.~\eqref{completeLO} reduce to
\begin{align}\label{asympLO}
V_1&=\frac{16\sqrt{2} \pi\alpha _s C_F s^2\Psi_{B_c^*}(0)\Psi_{J/\psi}(0)}{m_b^3 x^{\frac{3}{2}}}\nonumber\\
&=V_2=\frac{V_3}{2}=\frac{V_4}{4x}=A_1=A_2\nonumber\\
&=T_2=\frac{2s}{4s-1}T_3=\frac{s}{x(4s-1)}T_4=2 T_6\nonumber\\
&=s T'_2=T'_3=2T'_4.
\end{align}

At NLO, the complete analytical expressions of the form factors are too lengthy to be presented in this paper, while their asymptotic forms in the hierarchical heavy quark limit are given in the appendix~\ref{appendix123}.
Note that the form factor relations in Eqs.~\eqref{completeLO}, \eqref{asympLO}, and the appendix~\ref{appendix123} do not apply to the complete NLO results.

\section{Numerical results}\label{NUMERICALRESULTS}

To calculate the numerical results of the form factors and perform phenomenological analysis, we take $m_b = 4.75 \, \text{GeV}$, $m_c = 1.5 \, \text{GeV}$, and use 
the package \texttt{RunDec}'s function \texttt{AsRunDec} to compute the QCD running coupling at two-loop accuracy. Meanwhile, we adopt the following scheme for the number of active flavors:  $n_b = n_c = 1, n_l = 3$ when $\mu \geq 4.75 \, \text{GeV}$, $n_b = 0, n_c = 1, n_l = 3$ when $1.5 \, \text{GeV} \leq \mu < 4.75 \, \text{GeV}$, and $n_b = 0, n_c = 0, n_l = 3$ when $\mu < 1.5 \, \text{GeV}$.

\begin{figure}[!htbp]
	\centering
	\includegraphics[width=0.45\textwidth]{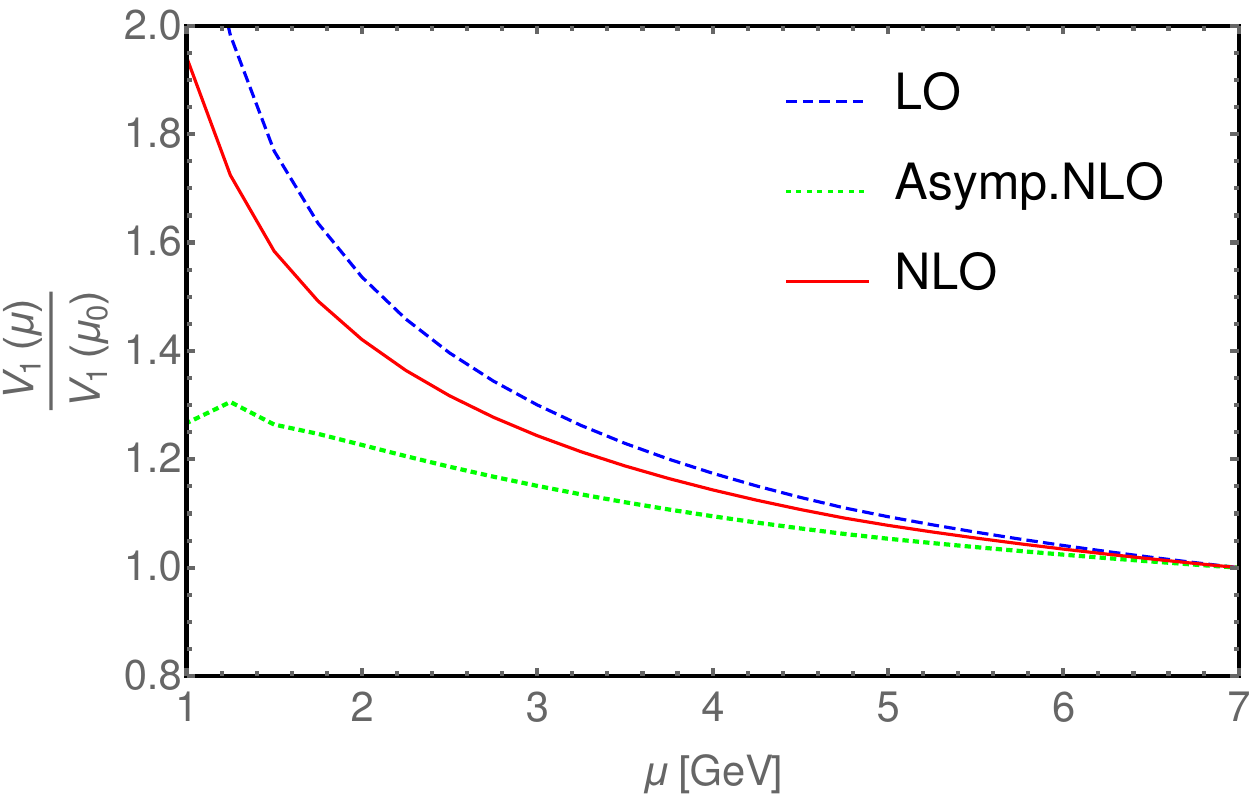}
	\caption{The renormalization scale $\mu$ dependence of the form factor ratio $V_1(\mu)/V_1(\mu_0)$ at LO, asymptotic NLO and complete NLO at the maximum recoil point $q^2=0$, where $\mu_0=7\,\text{GeV}$ and  $V_1^\text{LO}(\mu_0):V_1^\text{Asymp.NLO}(\mu_0):V_1^\text{NLO}(\mu_0)\approx 0.600: 0.828: 1$.}
	\label{fig:V1mu}
\end{figure}

\begin{figure}[!htbp]
	\centering
	\includegraphics[width=0.45\textwidth]{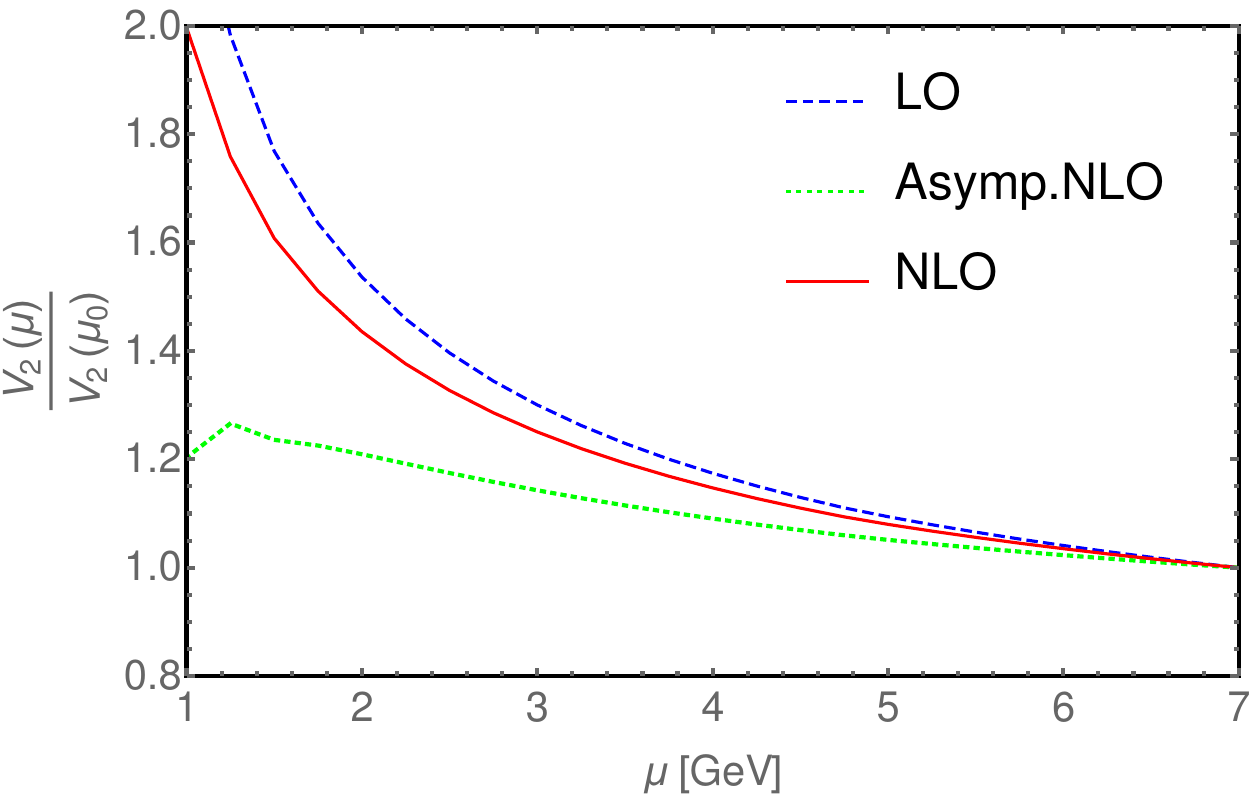}
	\caption{The same as Fig.~\ref{fig:V1mu}, but for  $V_2$ with $V_2^\text{LO}(\mu_0):V_2^\text{Asymp.NLO}(\mu_0):V_2^\text{NLO}(\mu_0)\approx 0.591: 0.803: 1$.}
	\label{fig:V2mu}
\end{figure}

\begin{figure}[!htbp]
	\centering
	\includegraphics[width=0.45\textwidth]{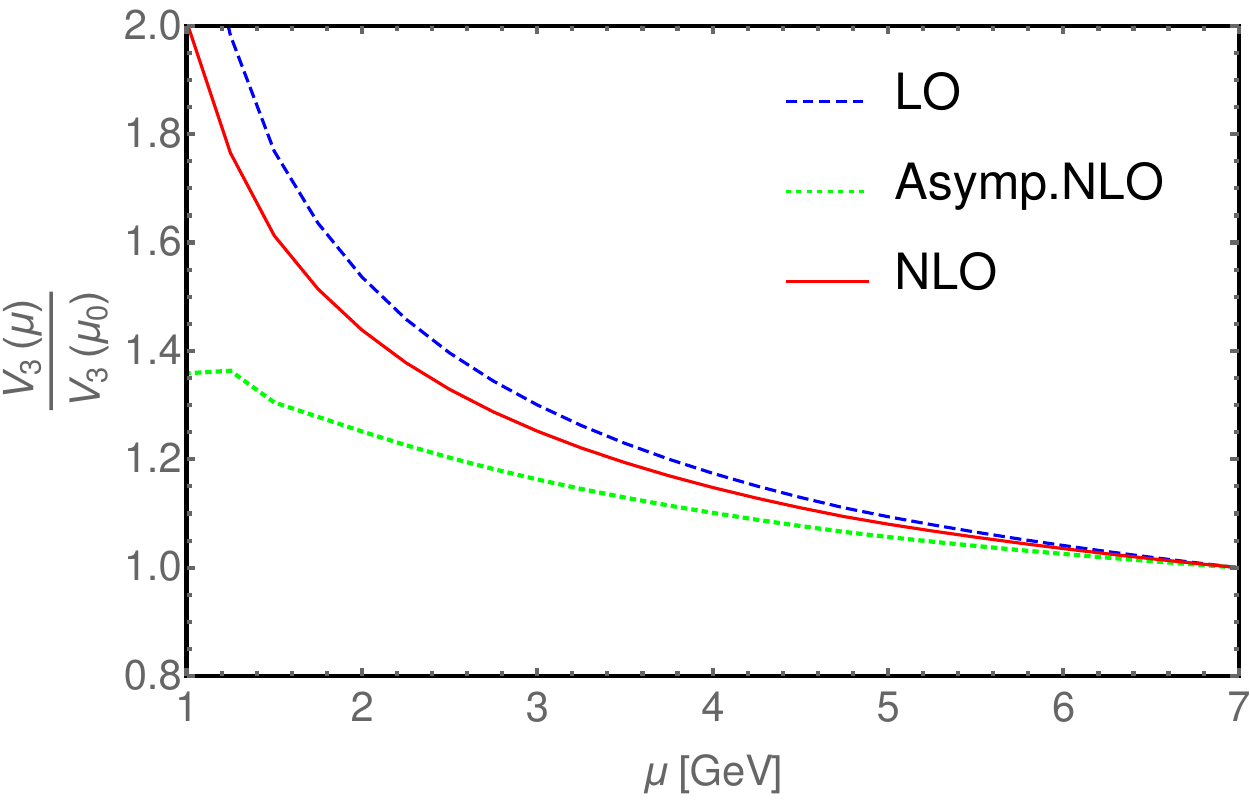}
	\caption{The same as Fig.~\ref{fig:V1mu}, but for  $V_3$ with $V_3^\text{LO}(\mu_0):V_3^\text{Asymp.NLO}(\mu_0):V_3^\text{NLO}(\mu_0)\approx 0.589: 0.831: 1$.}
	\label{fig:V3mu}
\end{figure}

\begin{figure}[!htbp]
	\centering
	\includegraphics[width=0.45\textwidth]{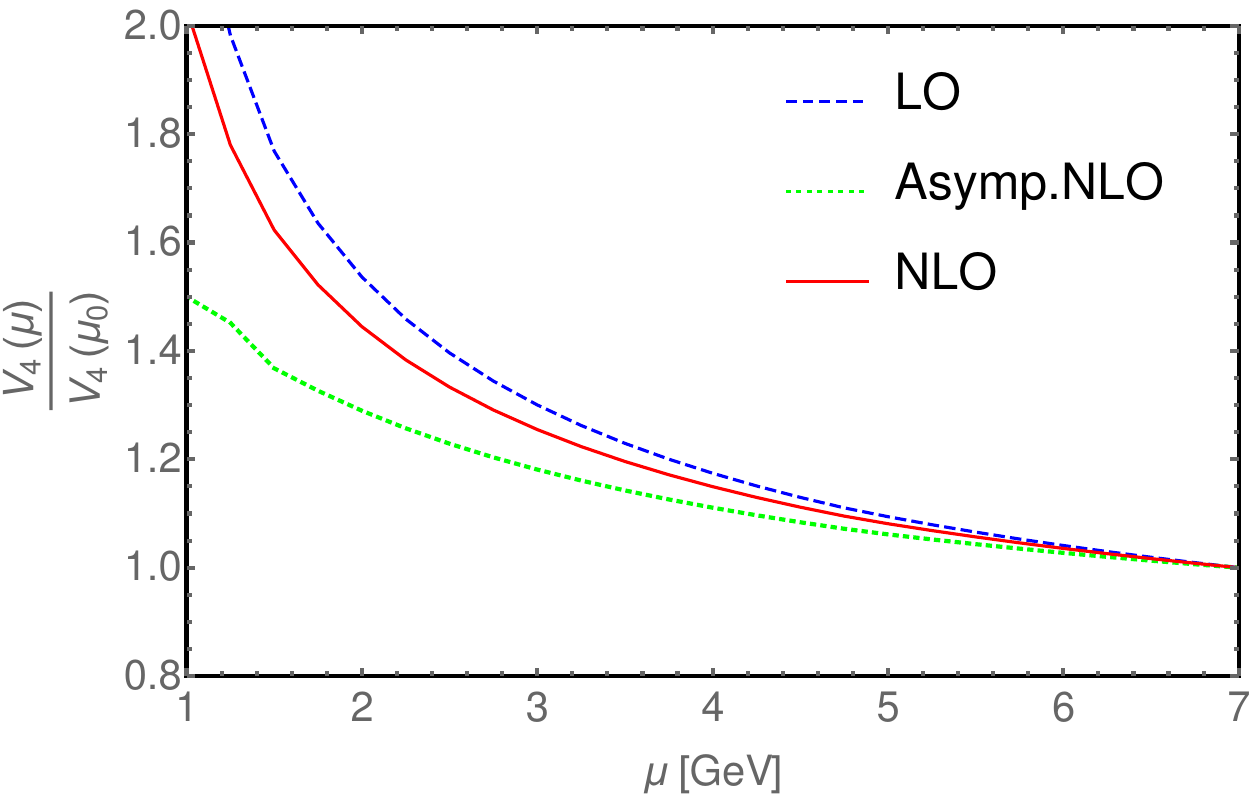}
	\caption{The same as Fig.~\ref{fig:V1mu}, but for  $V_4$ with $V_4^\text{LO}(\mu_0):V_4^\text{Asymp.NLO}(\mu_0):V_4^\text{NLO}(\mu_0)\approx 0.585: 0.855: 1$.}
	\label{fig:V4mu}
\end{figure}

\begin{figure}[!htbp]
	\centering
	\includegraphics[width=0.45\textwidth]{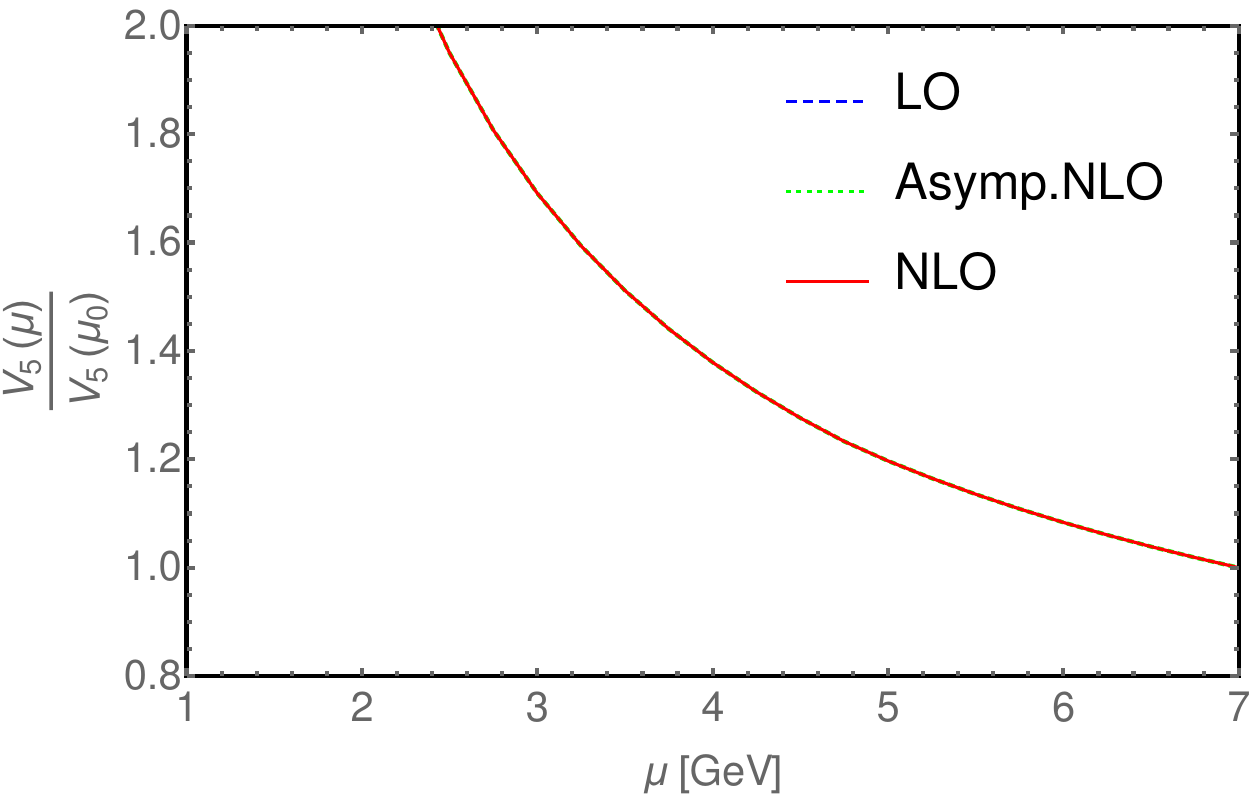}
	\caption{The same as Fig.~\ref{fig:V1mu}, but for  $V_5$ with $V_5^\text{LO}(\mu_0):V_5^\text{Asymp.NLO}(\mu_0):V_5^\text{NLO}(\mu_0)\approx 0: 0.168: 1$.}
	\label{fig:V5mu}
\end{figure}

\begin{figure}[!htbp]
	\centering
	\includegraphics[width=0.45\textwidth]{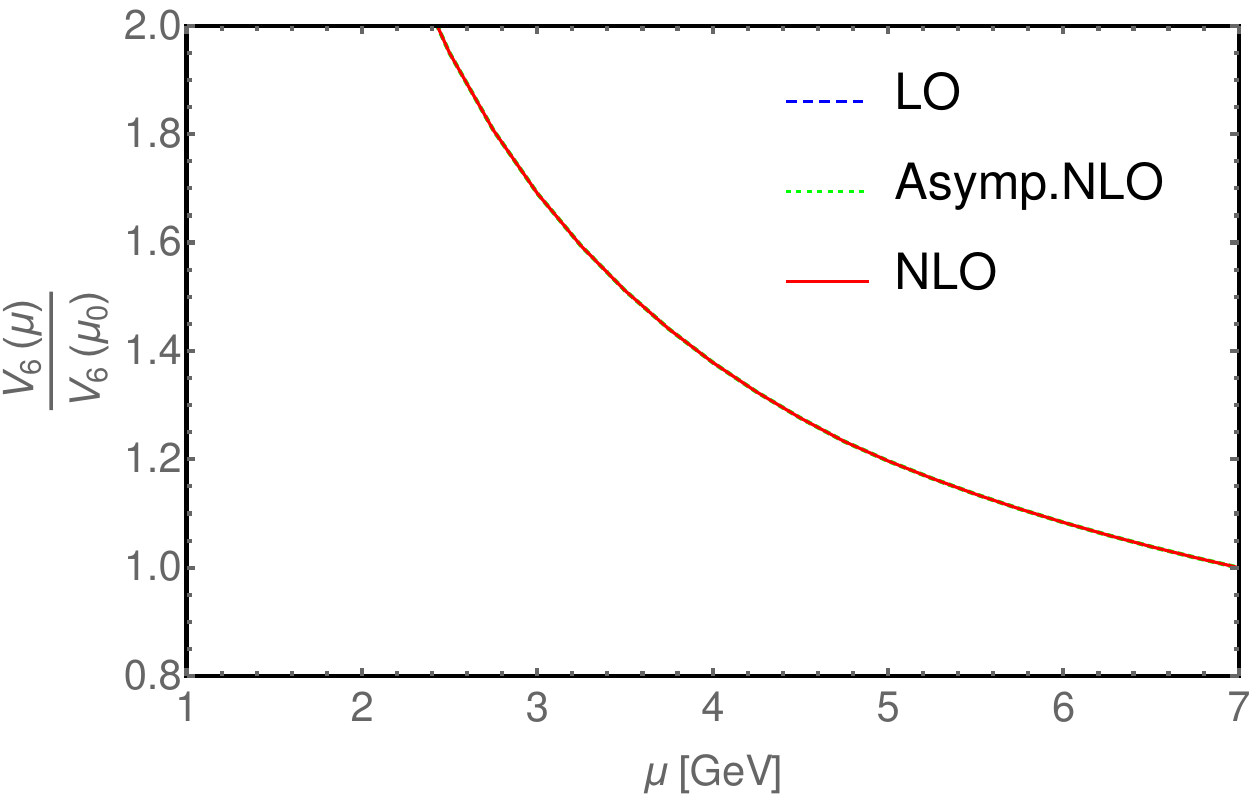}
	\caption{The same as Fig.~\ref{fig:V1mu}, but for  $V_6$ with $V_6^\text{LO}(\mu_0):V_6^\text{Asymp.NLO}(\mu_0):V_6^\text{NLO}(\mu_0)\approx 0: 0.168: 1$.}
	\label{fig:V6mu}
\end{figure}

\begin{figure}[!htbp]
	\centering
	\includegraphics[width=0.45\textwidth]{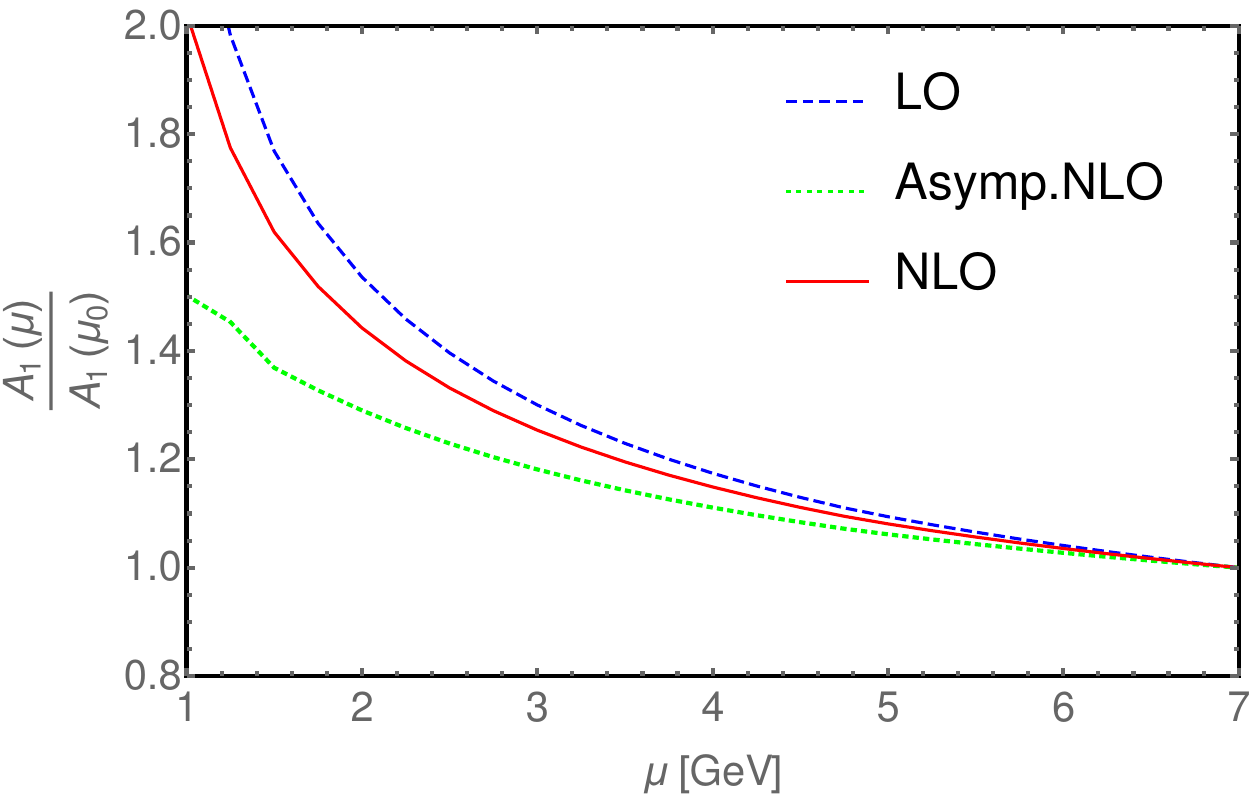}
	\caption{The same as Fig.~\ref{fig:V1mu}, but for  $A_1$ with $A_1^\text{LO}(\mu_0):A_1^\text{Asymp.NLO}(\mu_0):A_1^\text{NLO}(\mu_0)\approx 0.586: 0.858: 1$.}
	\label{fig:A1mu}
\end{figure}

\begin{figure}[!htbp]
	\centering
	\includegraphics[width=0.45\textwidth]{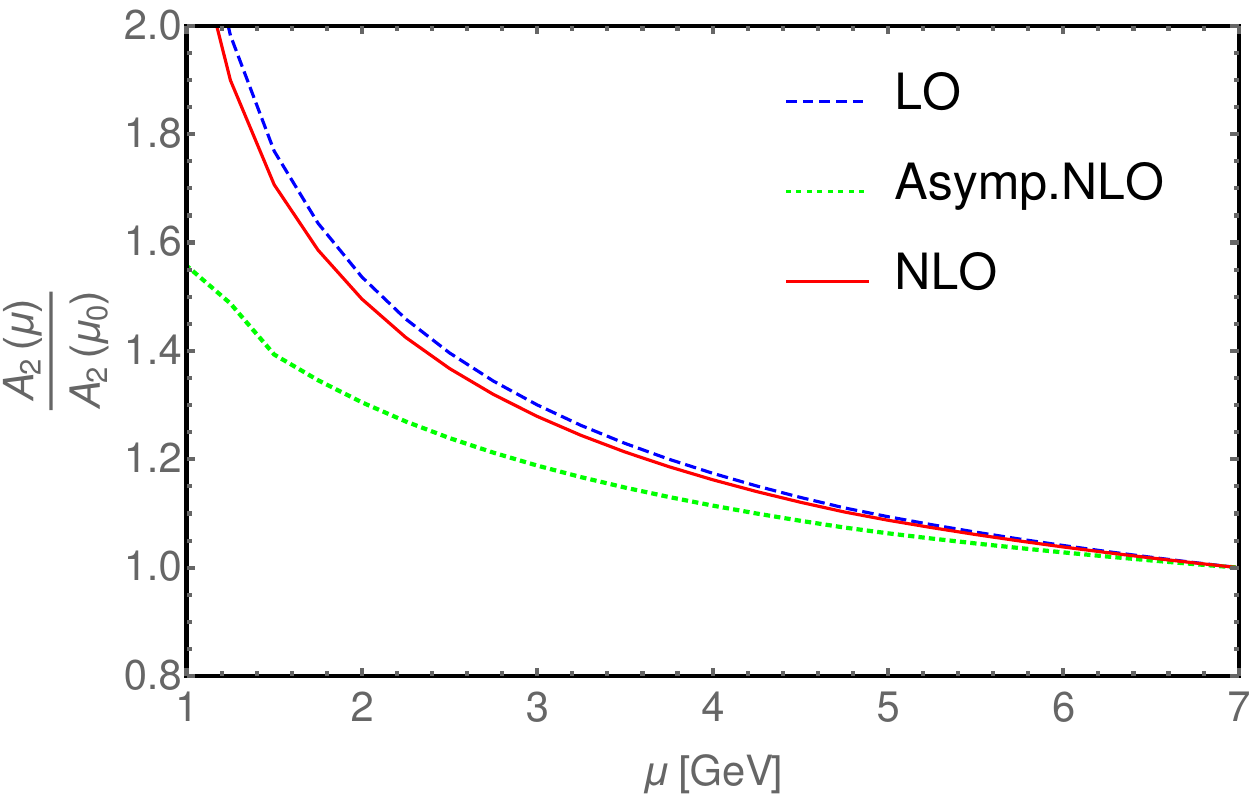}
	\caption{The same as Fig.~\ref{fig:V1mu}, but for  $A_2$ with $A_2^\text{LO}(\mu_0):A_2^\text{Asymp.NLO}(\mu_0):A_2^\text{NLO}(\mu_0)\approx 0.552: 0.819: 1$.}
	\label{fig:A2mu}
\end{figure}

\begin{figure}[!htbp]
	\centering
	\includegraphics[width=0.45\textwidth]{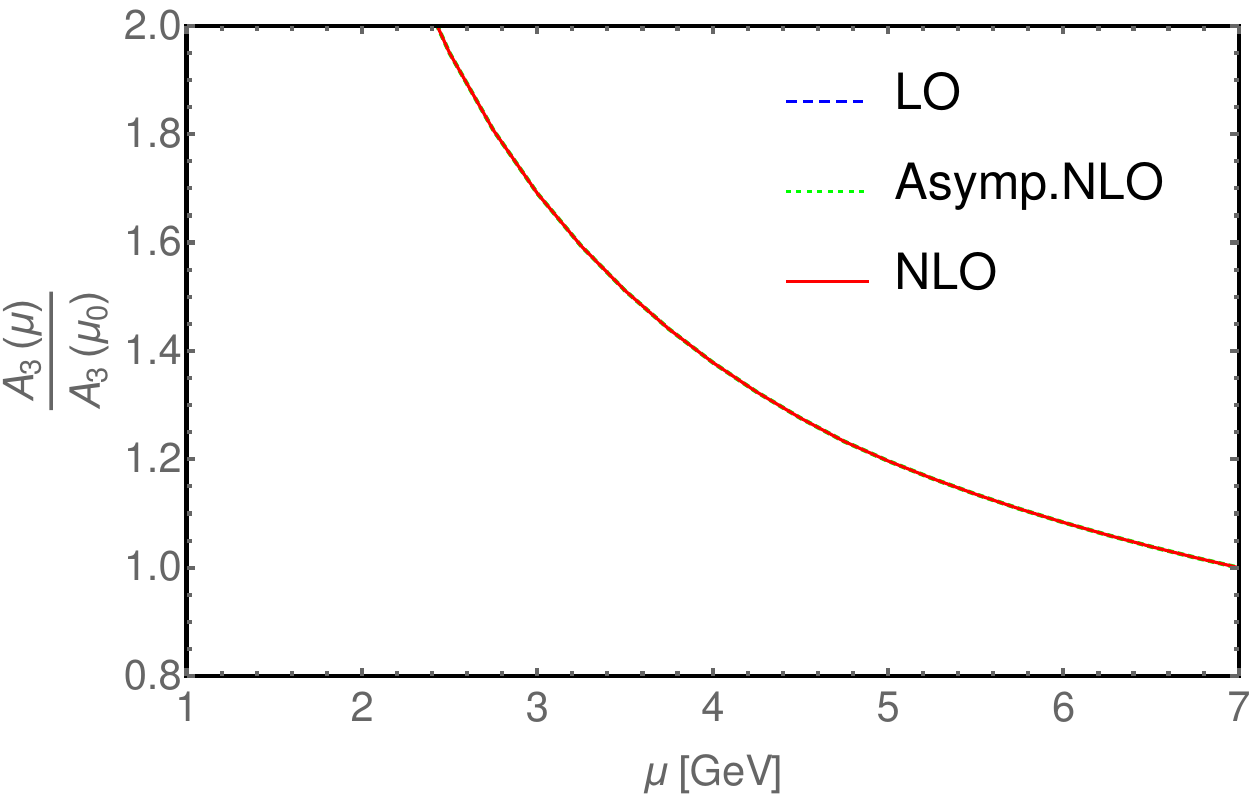}
	\caption{The same as Fig.~\ref{fig:V1mu}, but for  $A_3$ with $A_3^\text{LO}(\mu_0):A_3^\text{Asymp.NLO}(\mu_0):A_3^\text{NLO}(\mu_0)\approx 0: 0.077: 1$.}
	\label{fig:A3mu}
\end{figure}

\begin{figure}[!htbp]
	\centering
	\includegraphics[width=0.45\textwidth]{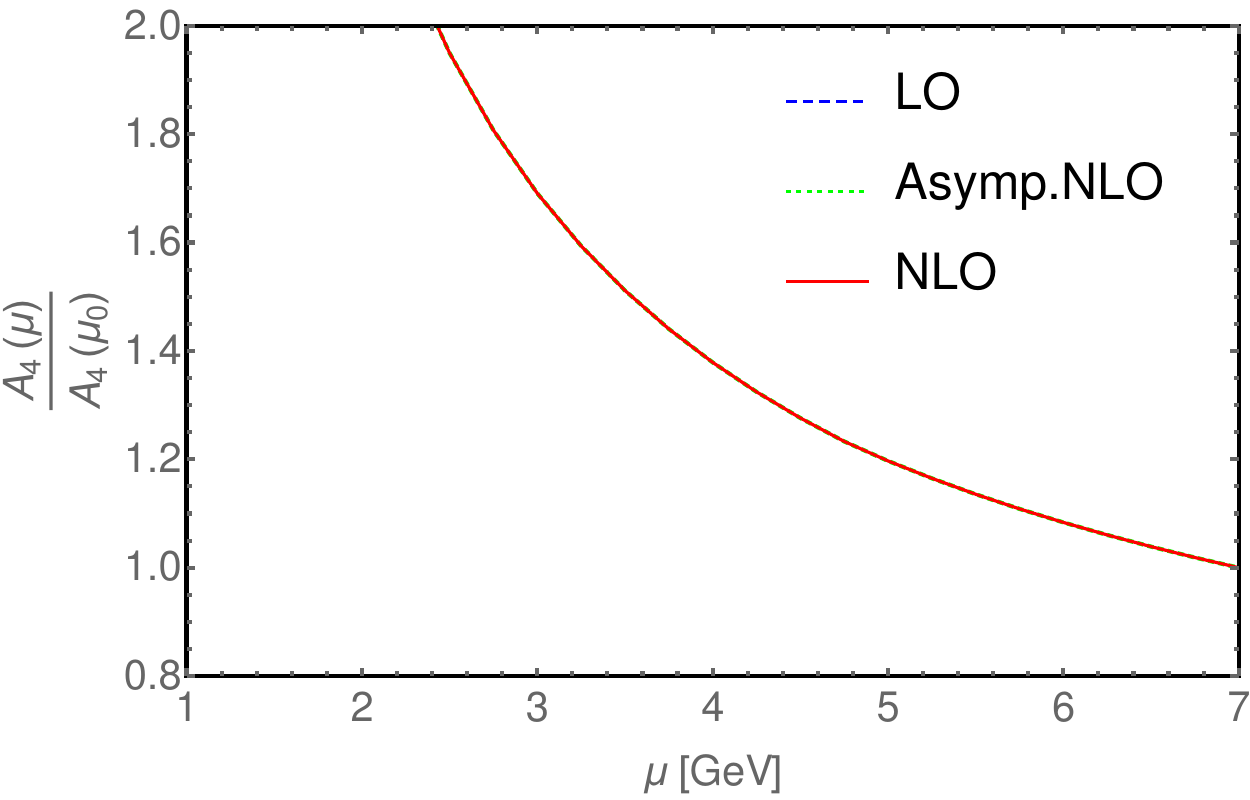}
	\caption{The same as Fig.~\ref{fig:V1mu}, but for  $A_4$ with $A_4^\text{LO}(\mu_0):A_4^\text{Asymp.NLO}(\mu_0):A_4^\text{NLO}(\mu_0)\approx 0: 0.266: 1$.}
	\label{fig:A4mu}
\end{figure}

\begin{figure}[!htbp]
	\centering
	\includegraphics[width=0.45\textwidth]{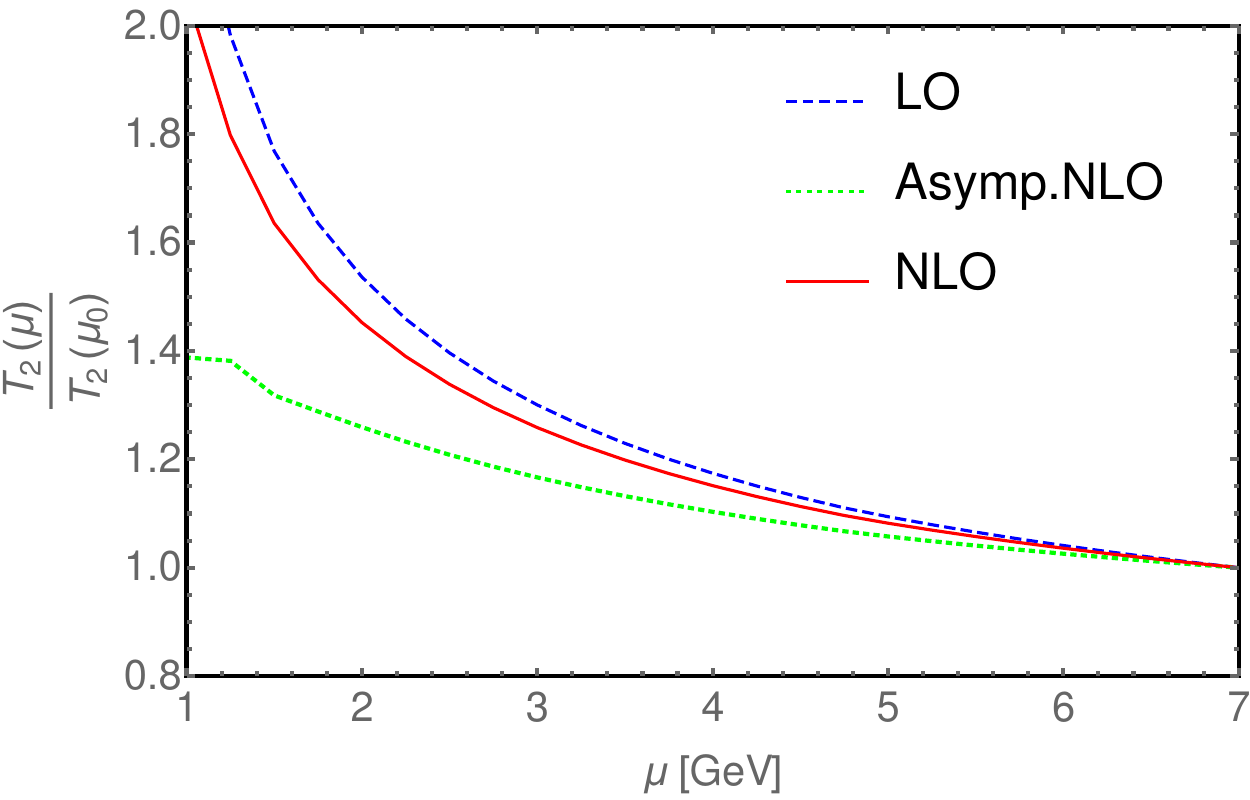}
	\caption{The same as Fig.~\ref{fig:V1mu}, but for  $T_2$ with $T_2^\text{LO}(\mu_0):T_2^\text{Asymp.NLO}(\mu_0):T_2^\text{NLO}(\mu_0)\approx 0.580: 0.824: 1$.}
	\label{fig:T2mu}
\end{figure}

\begin{figure}[!htbp]
	\centering
	\includegraphics[width=0.45\textwidth]{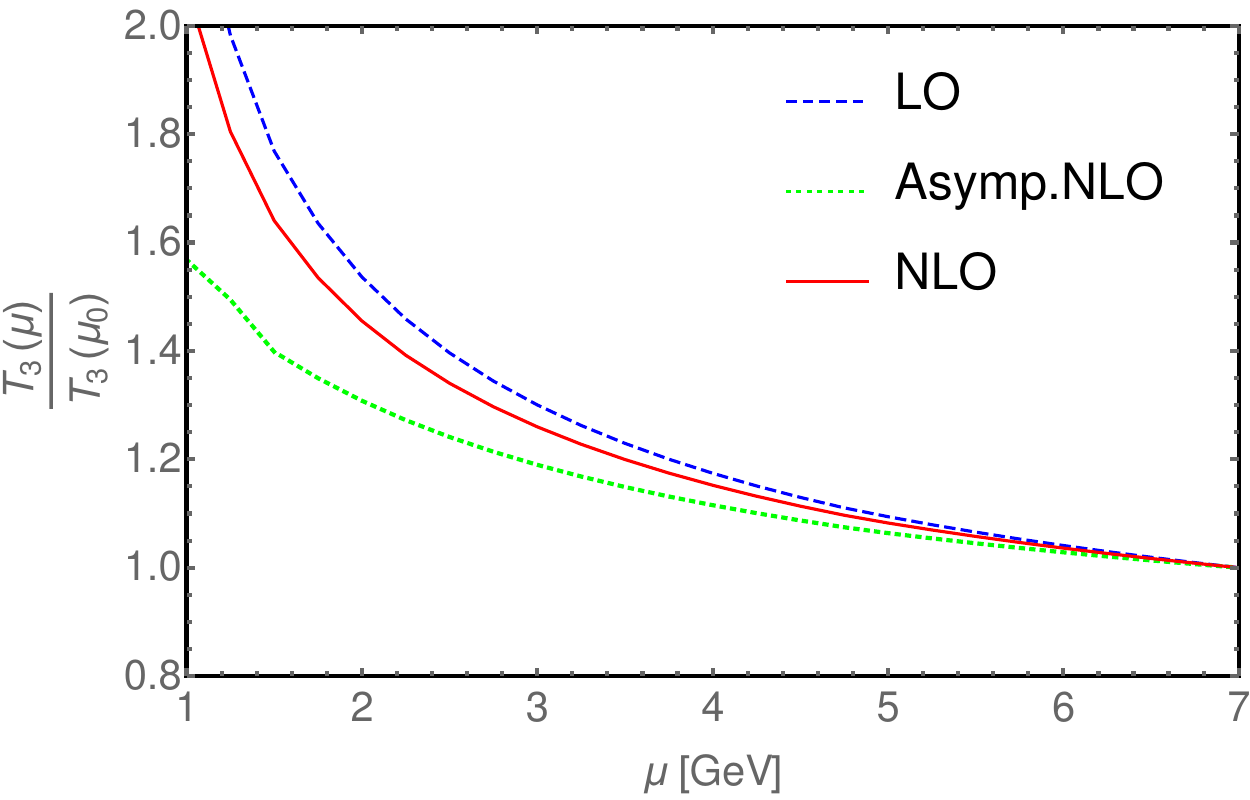}
	\caption{The same as Fig.~\ref{fig:V1mu}, but for  $T_3$ with $T_3^\text{LO}(\mu_0):T_3^\text{Asymp.NLO}(\mu_0):T_3^\text{NLO}(\mu_0)\approx 0.578: 0.860: 1$.}
	\label{fig:T3mu}
\end{figure}

\begin{figure}[!htbp]
	\centering
	\includegraphics[width=0.45\textwidth]{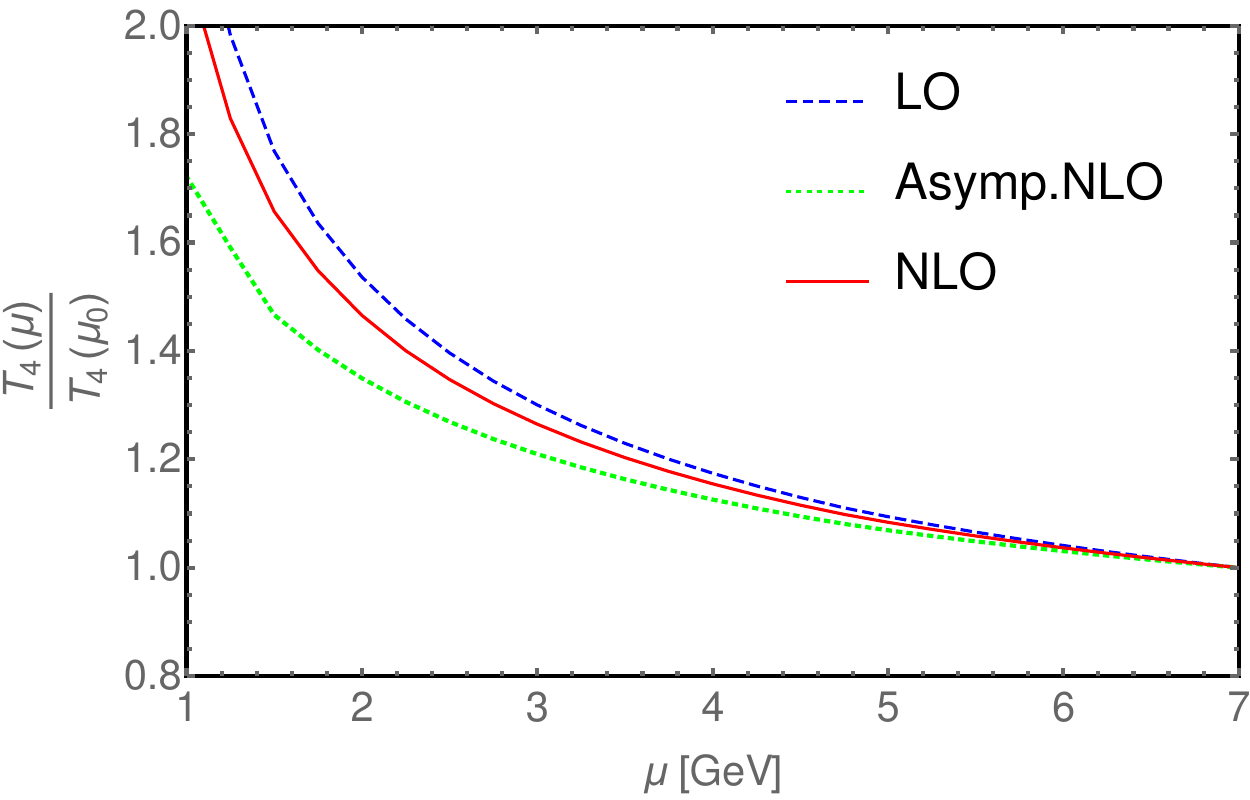}
	\caption{The same as Fig.~\ref{fig:V1mu}, but for  $T_4$ with $T_4^\text{LO}(\mu_0):T_4^\text{Asymp.NLO}(\mu_0):T_4^\text{NLO}(\mu_0)\approx 0.572: 0.885: 1$.}
	\label{fig:T4mu}
\end{figure}

\begin{figure}[!htbp]
	\centering
	\includegraphics[width=0.45\textwidth]{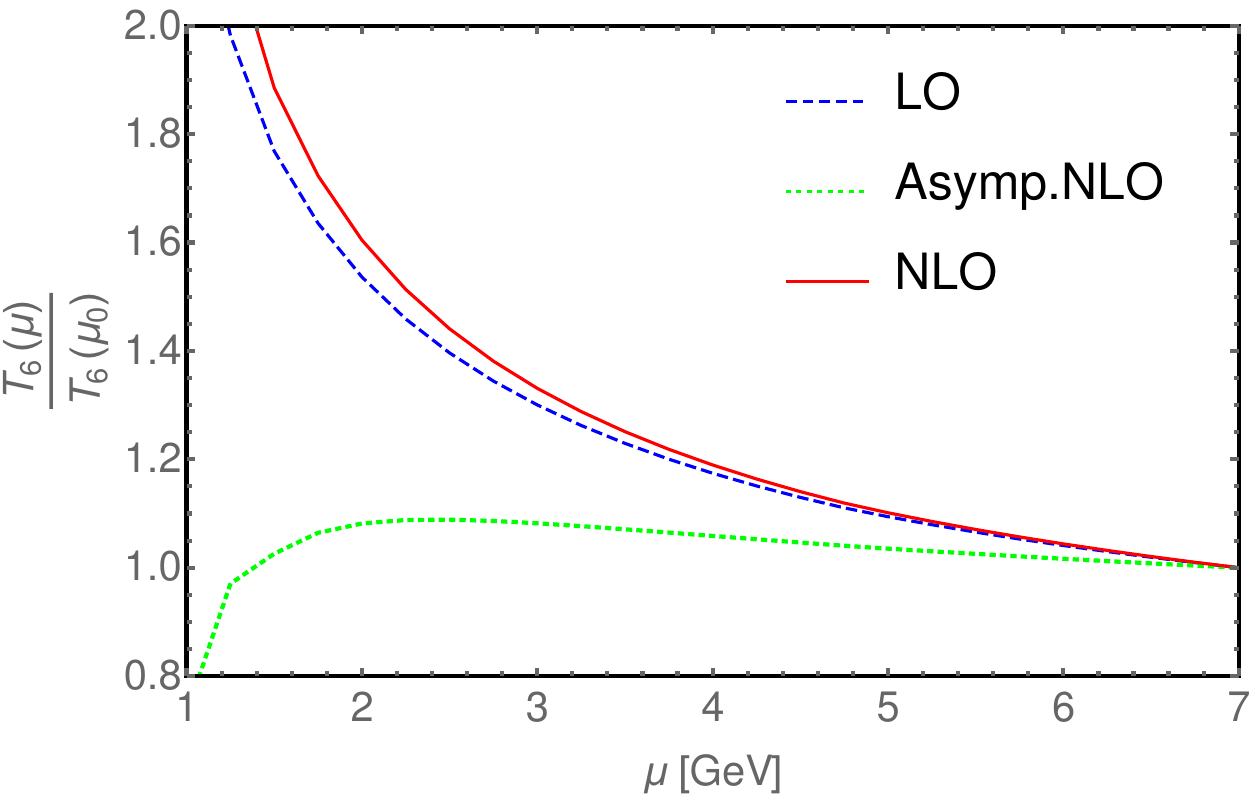}
	\caption{The same as Fig.~\ref{fig:V1mu}, but for  $T_6$ with $T_6^\text{LO}(\mu_0):T_6^\text{Asymp.NLO}(\mu_0):T_6^\text{NLO}(\mu_0)\approx 0.483: 0.591: 1$.}
	\label{fig:T6mu}
\end{figure}

\begin{figure}[!htbp]
	\centering
	\includegraphics[width=0.45\textwidth]{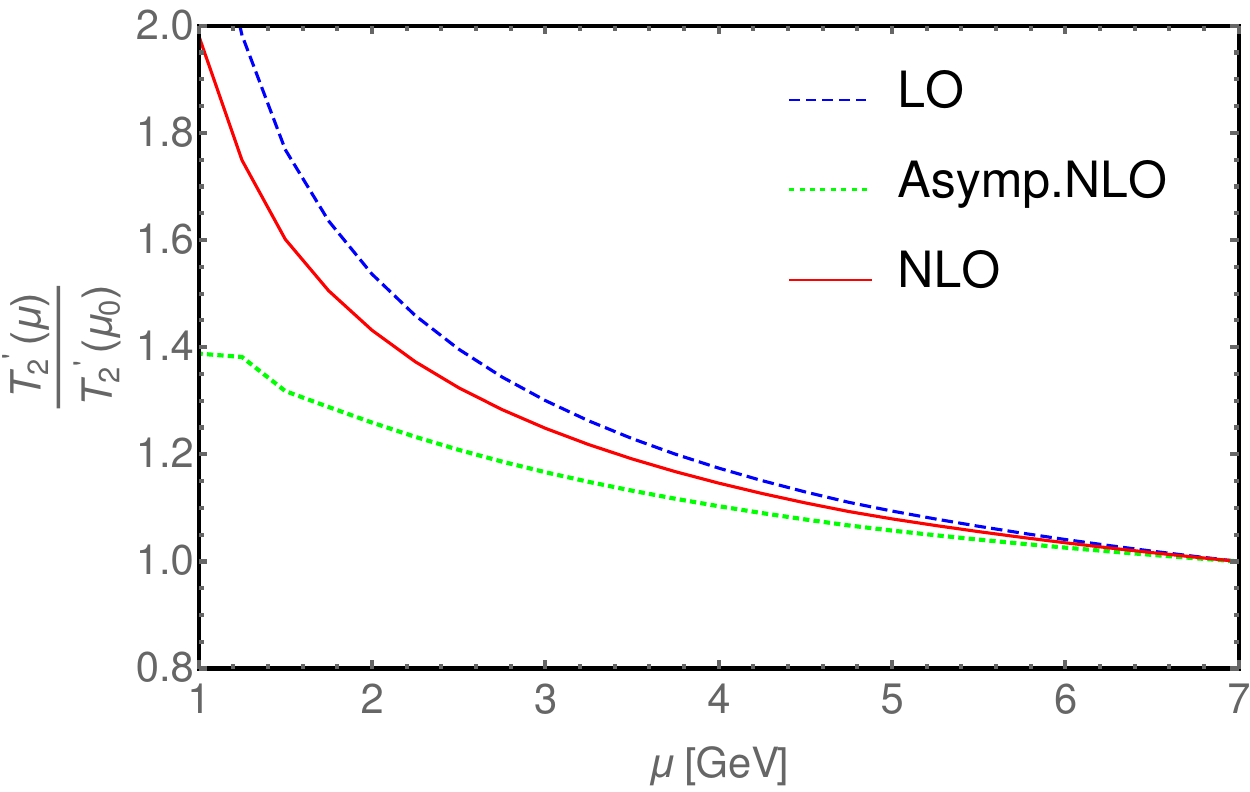}
	\caption{The same as Fig.~\ref{fig:V1mu}, but for  $T'_2$ with ${T'_2}^\text{LO}(\mu_0):{T'_2}^\text{Asymp.NLO}(\mu_0):{T'_2}^\text{NLO}(\mu_0)\approx 0.593: 0.843: 1$.}
	\label{fig:AT2mu}
\end{figure}

\begin{figure}[!htbp]
	\centering
	\includegraphics[width=0.45\textwidth]{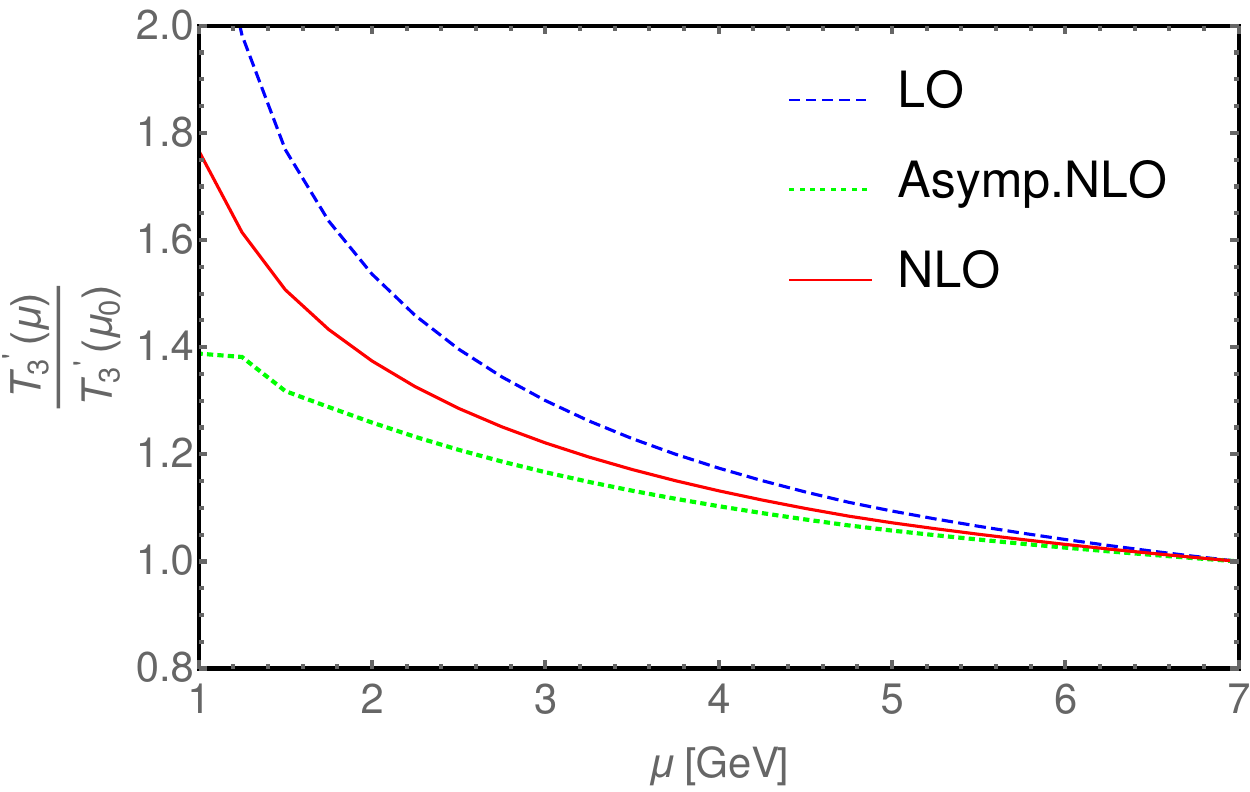}
	\caption{The same as Fig.~\ref{fig:V1mu}, but for  $T'_3$ with ${T'_3}^\text{LO}(\mu_0):{T'_3}^\text{Asymp.NLO}(\mu_0):{T'_3}^\text{NLO}(\mu_0)\approx 0.630: 0.895: 1$.}
	\label{fig:AT3mu}
\end{figure}


\begin{figure}[!htbp]
	\centering
	\includegraphics[width=0.45\textwidth]{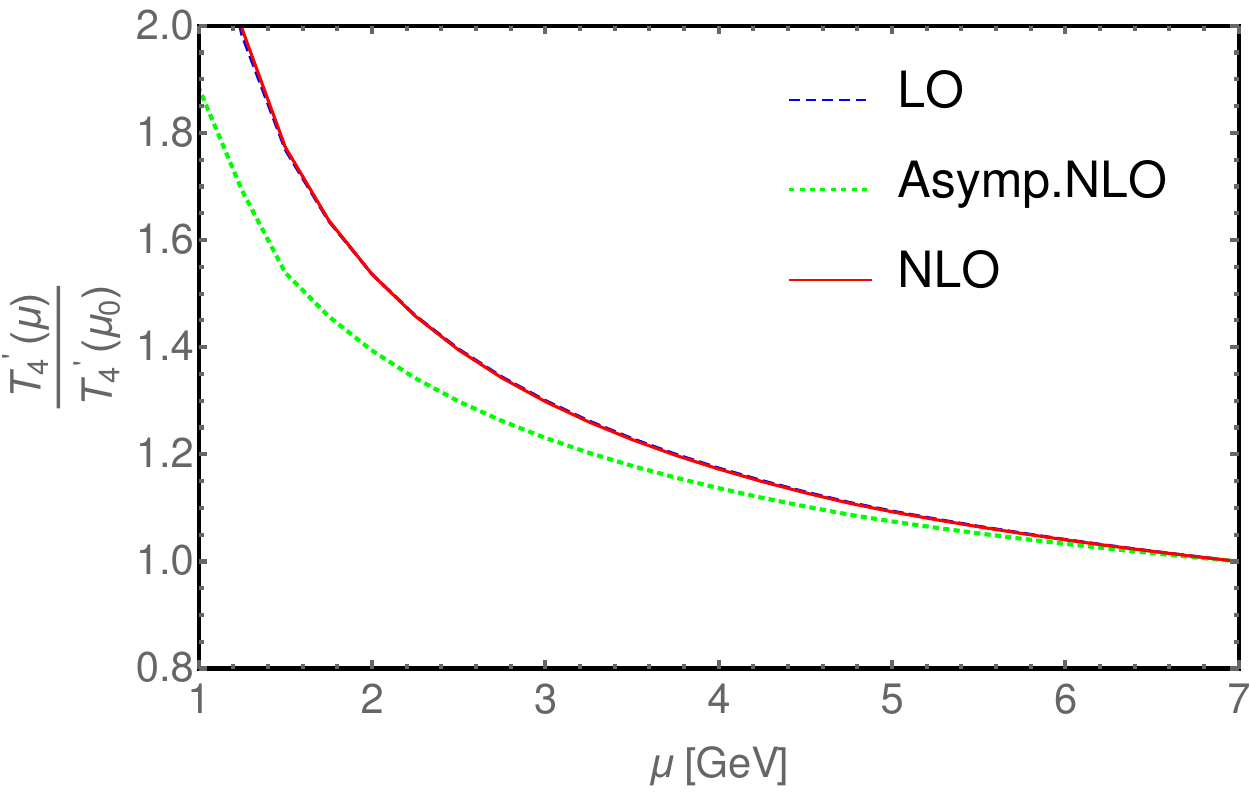}
	\caption{The same as Fig.~\ref{fig:V1mu}, but for  $T'_4$ with ${T'_4}^\text{LO}(\mu_0):{T'_4}^\text{Asymp.NLO}(\mu_0):{T'_4}^\text{NLO}(\mu_0)\approx 0.527: 0.852: 1$.}
	\label{fig:AT4mu}
\end{figure}

\clearpage

In Figs.~\ref{fig:V1mu}--\ref{fig:AT4mu}, we present the renormalization scale $\mu$ dependence of the form factor ratio $\frac{F_i(\mu)}{F_i(\mu_0)}$ at LO, asymptotic NLO and complete NLO accuracy at the maximum recoil point $q^2=0$. Here, $\mu_0=7\,\text{GeV}$, and the wave functions at the origin have been canceled from the ratio.
From Figs.~\ref{fig:V1mu}--\ref{fig:AT4mu}, one can see that, in general, the NLO corrections to the form factors have reduced the renormalization scale dependence. This can be explained by the asymptotic expressions of the NLO form factors provided in the appendix~\ref{appendix123}. It can be verified that the form factor expressions satisfy the renormalization group invariance, which implies that the scale dependence of the LO form factors is at ${\cal O}(\alpha_s^2)$, while that of the NLO form factors is at ${\cal O}(\alpha_s^3)$.
We also observe that for the form factors $T_6$ and $T'_4$, the NLO corrections do not reduce the renormalization scale dependence, which is attributed to the large $\alpha_s^2$ terms in the NLO corrections.


\begin{figure}[!htbp]
	\centering
	\includegraphics[width=0.45\textwidth]{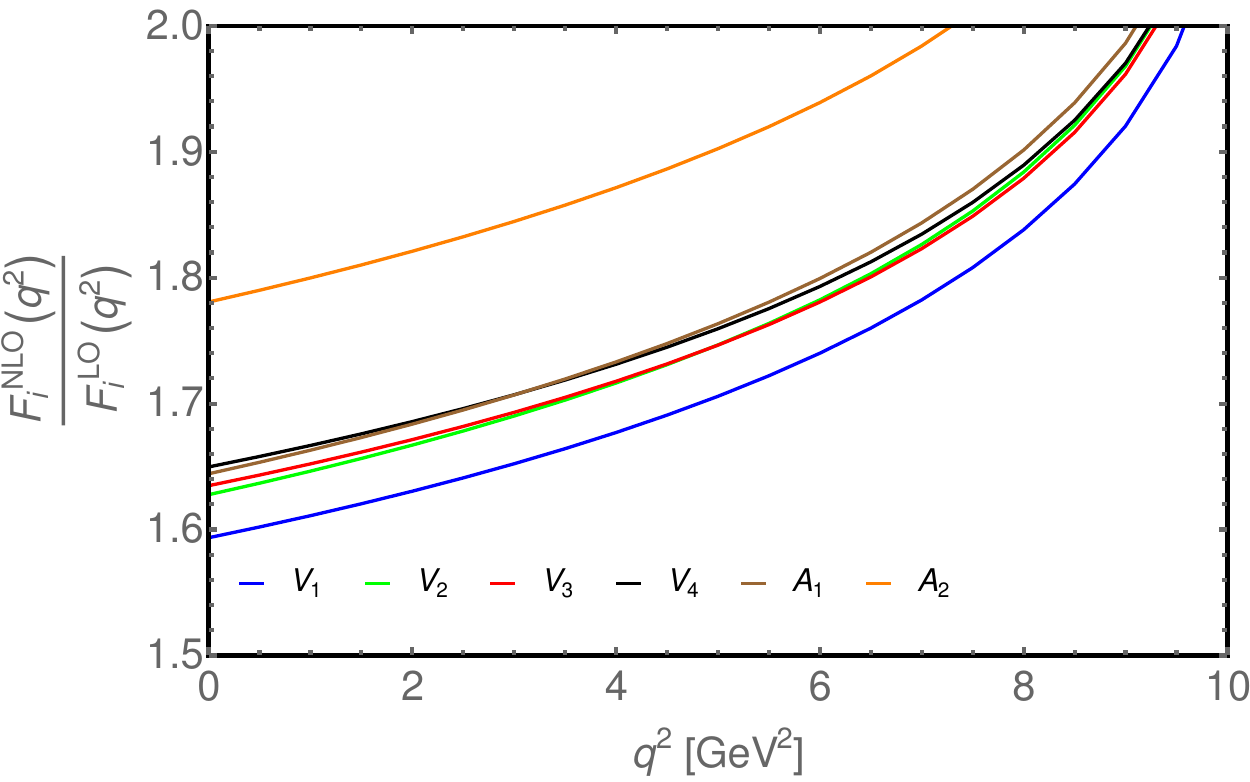}\\
	\includegraphics[width=0.45\textwidth]{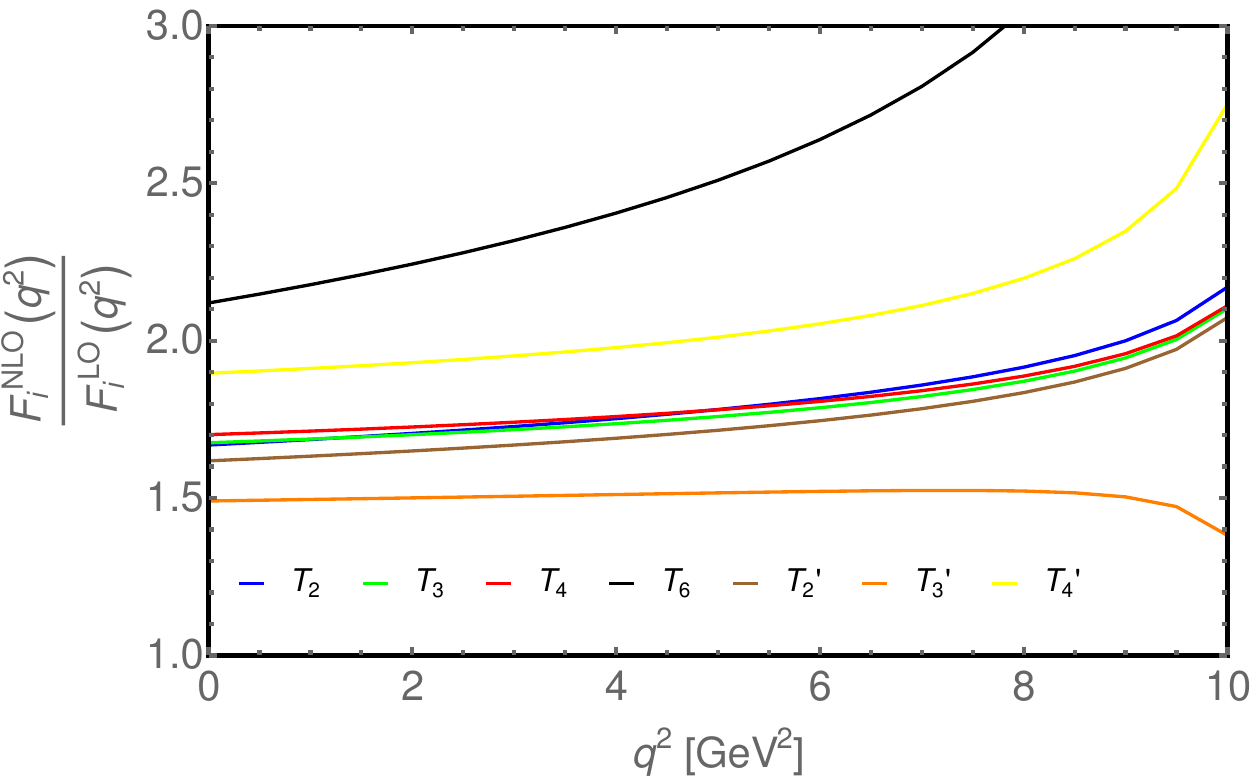}
	\caption{The $q^2$ dependence of $F_i^\text{NLO}(q^2)/F_i^\text{LO}(q^2)$ with  the form factor $F_i$ standing for $V_{1,2,3,4}$, $A_{1,2}$, $T_{2,3,4,6}$ and $T'_{2,3,4}$. In the computation, the renormalization scale is fixed at $\mu=3\,\text{GeV}$. }
	\label{fig:Fiq2}
\end{figure}

To further investigate the convergence behaviors of the NLO corrections to the form factors at different $q^2$ values, we plot the squared transfer momentum $q^2$ dependence for the ratios of NLO to LO form factors, $F_i^\text{NLO}(q^2)/F_i^\text{LO}(q^2)$, in Fig.~\ref{fig:Fiq2}.
The figure shows that the NLO corrections to the form factors are both significant and convergent in the maximum recoil point and relatively small $q^2$ region. However, at large $q^2$, the convergence breaks down, which means that the NLO calculations within  NRQCD  for the form factors may not be applicable at large $q^2$ values~\cite{Qiao:2011yz}.

%

\begin{table}[!htbp]
\begin{center}
	\caption{The NRQCD + Lattice predictions for the $B_c^*\to J/\psi$ form factors at the maximum recoil point $q^2=0$, where the first uncertainties come from  choosing the renormalization scale as $\mu=3_{-1.5}^{+4}\text{GeV}$ in the NRQCD calculation and the second uncertainties come from  the lattice QCD data errors. For comparison, we also list the corresponding LFQM results. }
	\label{tab:nrlatt}
	\setlength{\tabcolsep}{2.0mm}
	\renewcommand{\arraystretch}{1.6}
		\begin{tabular}{c|c|c}
			\hline
			{$$}
			& $\text{NRQCD+Lattice}$
			& $\text{LFQM}$~\cite{Chang:2019obq,Chang:2020xvu}
   \\
			\hline
			\multirow{1}*{$V_1$}
			& \multirow{1}*{$0.4320_{-0.0048}^{+0.0030}\pm 0.0448$}
			& \multirow{1}*{$0.56_{-0.01-0.17}^{+0.01+0.17}$}
              \\
			\hline
			{$	V_2 $}
			& {$0.2295_{-0.0004}^{+0.0003}\pm{0.0238}$}
			& {$0.33_{-0.01-0.04}^{+0.01+0.05}$}
	            \\
			\hline
			{$V_3$}
			& {$0.8865_{+0.0003}^{+0.0001}\pm{0.0919}$}
			& {$1.17_{-0.02-0.29}^{+0.02+0.23} $}
		                \\
			\hline
			$V_4$
			& {$0.4294_{+0.0018}^{-0.0009}\pm{0.0445}$}
			& {$0.65_{-0.01-0.19}^{+0.01+0.20}$}
		               \\
			\hline
			{$V_5 $}
			& {$0.1303_{+0.0569}^{-0.0338}\pm{0.0135}$}
			& {$0.20_{-0.00-0.02}^{+0.00+0.02}$}
		               \\
			\hline
			{$V_6$}
			& {$0.1303_{+0.0569}^{-0.0338}\pm{0.0135}$}
			& {$0.20_{-0.00-0.02}^{+0.00+0.02} $}
		               \\
			\hline
			{$A_1$}
			& {$0.4458_{+0.0013}^{-0.0006}\pm{0.0462}$}
			& {$0.54_{-0.01-0.17}^{+0.01+0.16}$}
		                \\
			\hline
			{$A_2$}
			& {$0.2510_{+0.0090}^{-0.0053}\pm{0.0260}$}
			& {$0.35_{-0.00}^{+0.00}$}
		               \\
			\hline
			{$A_3$}
			& {$0.0942_{+0.0411}^{-0.0244}\pm{0.0098}$}
			& {$0.13_{-0.00-0.02}^{+0.00+0.03}$}
		               \\
			\hline
			{$A_4$}
			& {$0.1092_{+0.0477}^{-0.0284}\pm{0.0113}$}
			& {$0.14_{-0.00-0.02}^{+0.00+0.02}$}
		                \\
		     \hline
		     {$T_2$}
		     & {$0.2352_{+0.0021}^{-0.0012}\pm{0.0244}$}
		     & {$-$}
		     \\           
		   \hline
		   {$T_3$}
		   & {$0.5724_{+0.0062}^{-0.0035}\pm{0.0593}$}
		   & {$-$}
		   \\             
		    \hline
		    {$T_4$}
		    & {$0.2790_{+0.0049}^{-0.0028}\pm{0.0289}$}
		    & {$-$}
		    \\            
		      \hline
		      {$T_6$}
		      & {$0.2211_{+0.0221}^{-0.0131}\pm{0.0229}$}
		      & {$-$}
		      \\          
		     \hline
		     {$T'_2$}
		     & {$0.4387_{-0.0018}^{+0.0012}\pm{0.0455}$}
		     & {$-$}
		     \\           
		    \hline
		    {$T'_3$}
		    & {$0.4546_{-0.0190}^{+0.0115}\pm{0.0471}$}
		    & {$-$}
		    \\            
		   \hline
		   {$T'_4$}
		   & {$0.3805_{+0.0230}^{-0.0136}\pm{0.0394}$}
		   & {$-$}
		   \\              
			\hline
			\end{tabular}
\end{center}		
\end{table}

As following, we will proceed to make the theoretical predictions for the $B_c^*\to J/\psi$ form factors involving various currents.
To determine $B_c^*$  wave function at the origin, 
we  employ the heavy quark spin symmetry~\cite{Bodwin:1994jh,Tao:2023pzv}  and approximate
the spin-triplet wave function to the spin-singlet wave function:
\begin{align}
\Psi_{B_c^*}(0)\approx\Psi_{B_c}(0).
\end{align}
Then the product of $B_c$ and $J/\psi$ wave functions at the origin, $\Psi_{B_c}(0)\Psi_{J/\psi}(0)$, can be extracted by combining the NLO NRQCD results for the $B_c\to J/\psi$ form factors involving vector and axial-vector currents with the corresponding lattice QCD results~\cite{Tao:2022yur,Harrison:2020gvo}. 
In a word, we can obtain the NRQCD + Lattice results for the $B_c^*\to J/\psi$ form factors using the following formula:
\begin{align}\label{nrlattformu}
F_{i,\text{NRQCD+Lattice}}^{B_c^*\to J/\psi}(q^2)=\frac{1}{4}\sum_{j=1}^{4}\frac{F_{i,\text{NRQCD}}^{B_c^*\to J/\psi}(q^2)}{F_{j,\text{NRQCD}}^{B_c\to J/\psi}(q^2)} F_{j,\text{Lattice}}^{B_c\to J/\psi}(q^2)
\end{align}
where $F_{j}^{B_c\to J/\psi}\in\{V,A_{0,1,2}\}^{B_c\to J/\psi}$ (see Ref.~\cite{Harrison:2020gvo}), 
and in the calculation of the ratio, we  no longer perform a series expansion in $\alpha_s$.
Note that the values of $q^2$ in Eq.~\eqref{nrlattformu} can not be too large,
because the NRQCD results may not be applicable at large $q^2$.
In Table~\ref{tab:nrlatt}, we provide the NRQCD + Lattice predictions for the $B_c^*\to J/\psi$ form factors at the maximum recoil point $q^2=0$, along with the corresponding LFQM results~\cite{Chang:2019obq,Chang:2020xvu}.
As shown in the table, the NRQCD + Lattice predictions for the vector and axial-vector  form factors roughly agree with the LFQM results, but are slightly smaller.
Additionally, in the NRQCD + Lattice results, for most of the form factors, the uncertainties from the lattice data dominate over the uncertainties from the renormalization scale. However, $V_{5,6}$ and $A_{3,4}$ are exceptions, as their LO contributions are zero.

\begin{figure}[!htbp]
	\centering
	\includegraphics[width=0.45\textwidth]{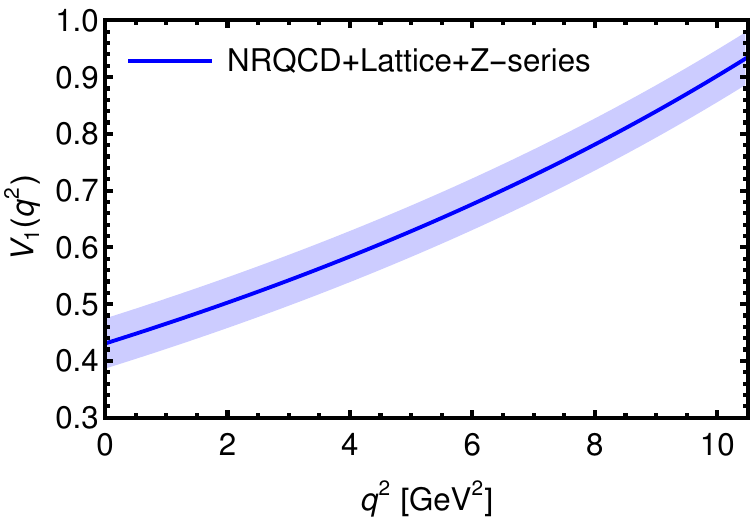}
	\caption{The NRQCD + Lattice + Z-series prediction for the $B_c^*\to J/\psi$ form factor $V_1(q^2)$ over the full physical $q^2$ range $0\leq q^2\leq (m_{B_c^*}-m_{J/\psi})^2$, where the error band stems from the uncertainties of the lattice QCD data.}
	\label{fig:V1q2}
\end{figure}

\begin{figure}[!htbp]
	\centering
	\includegraphics[width=0.45\textwidth]{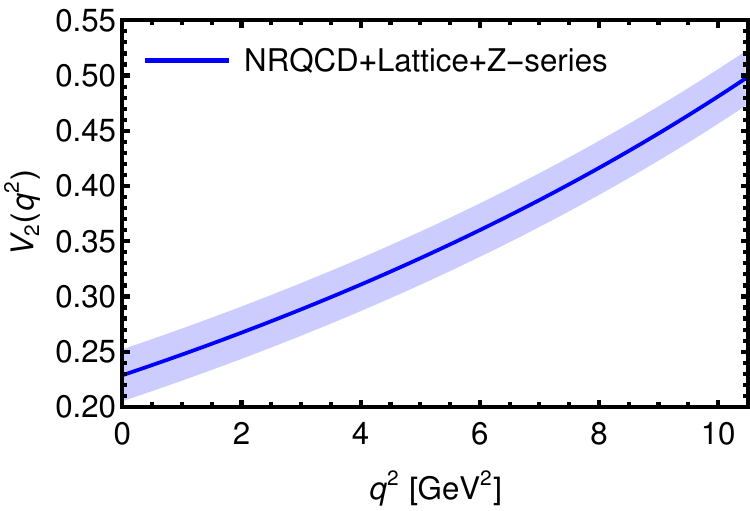}
	\caption{The same as Fig.~\ref{fig:V1q2}, but for $V_2(q^2)$.  }
	\label{fig:V2q2}
\end{figure}

\begin{figure}[!htbp]
	\centering
	\includegraphics[width=0.45\textwidth]{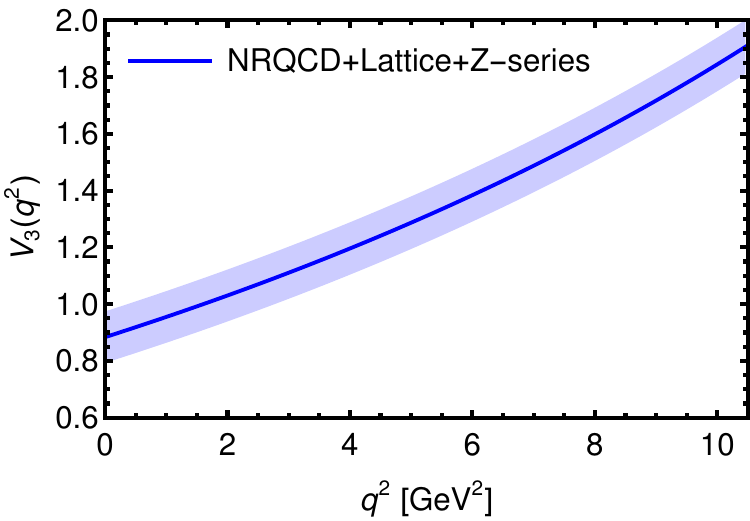}
	\caption{The same as Fig.~\ref{fig:V1q2}, but for $V_3(q^2)$.  }
	\label{fig:V3q2}
\end{figure}

\begin{figure}[!htbp]
	\centering
	\includegraphics[width=0.45\textwidth]{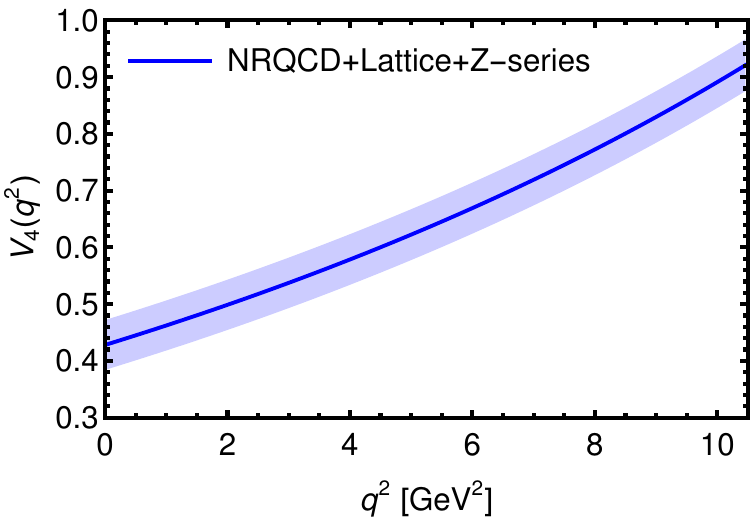}
	\caption{The same as Fig.~\ref{fig:V1q2}, but for $V_4(q^2)$.  }
	\label{fig:V4q2}
\end{figure}

\begin{figure}[!htbp]
	\centering
	\includegraphics[width=0.45\textwidth]{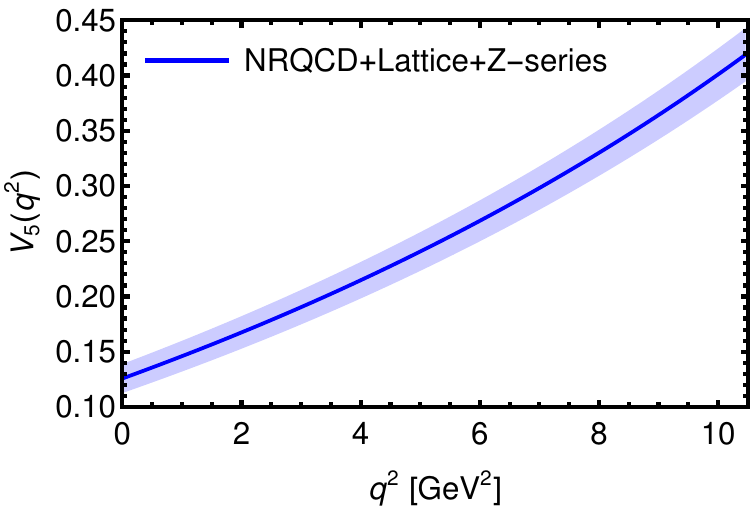}
	\caption{The same as Fig.~\ref{fig:V1q2}, but for $V_5(q^2)$.  }
	\label{fig:V5q2}
\end{figure}

\begin{figure}[!htbp]
	\centering
	\includegraphics[width=0.45\textwidth]{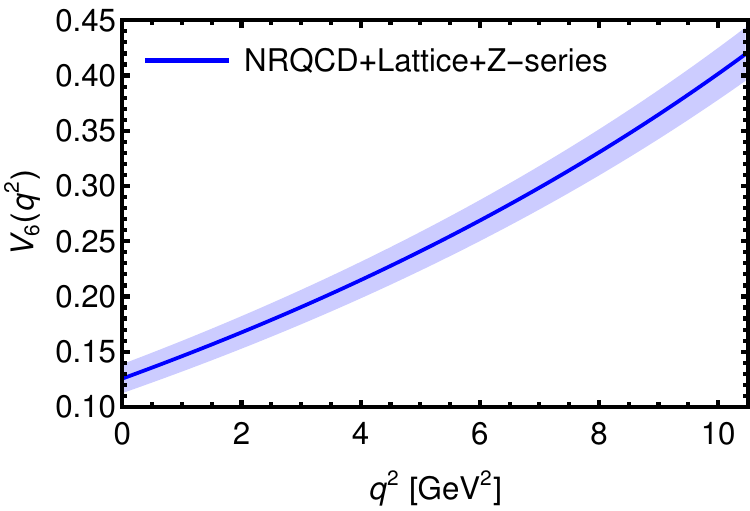}
	\caption{The same as Fig.~\ref{fig:V1q2}, but for $V_6(q^2)$.  }
	\label{fig:V6q2}
\end{figure}

\begin{figure}[!htbp]
	\centering
	\includegraphics[width=0.45\textwidth]{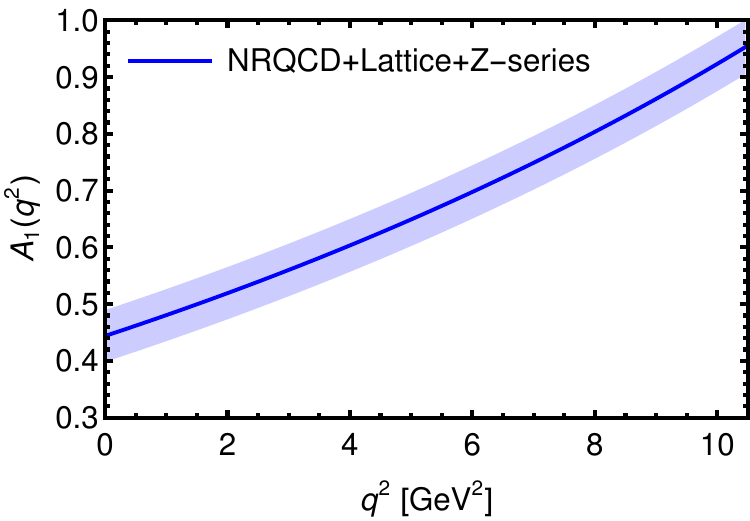}
	\caption{The same as Fig.~\ref{fig:V1q2}, but for $A_1(q^2)$.  }
	\label{fig:A1q2}
\end{figure}

\begin{figure}[!htbp]
	\centering
	\includegraphics[width=0.45\textwidth]{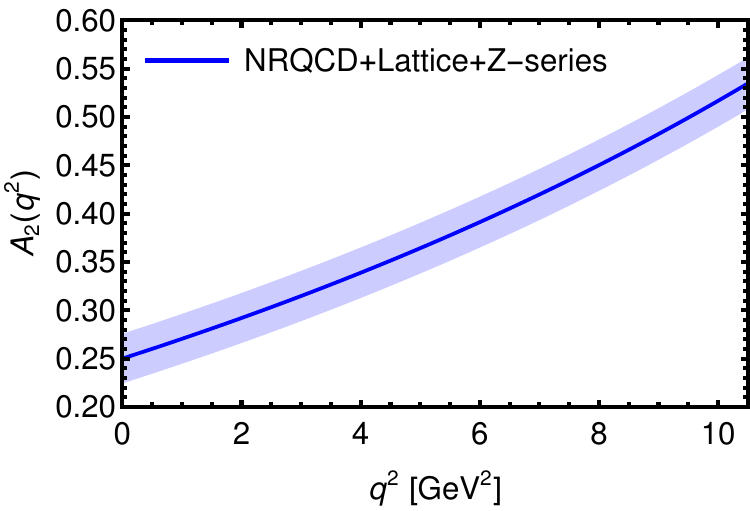}
	\caption{The same as Fig.~\ref{fig:V1q2}, but for $A_2(q^2)$.  }
	\label{fig:A2q2}
\end{figure}

\begin{figure}[!htbp]
	\centering
	\includegraphics[width=0.45\textwidth]{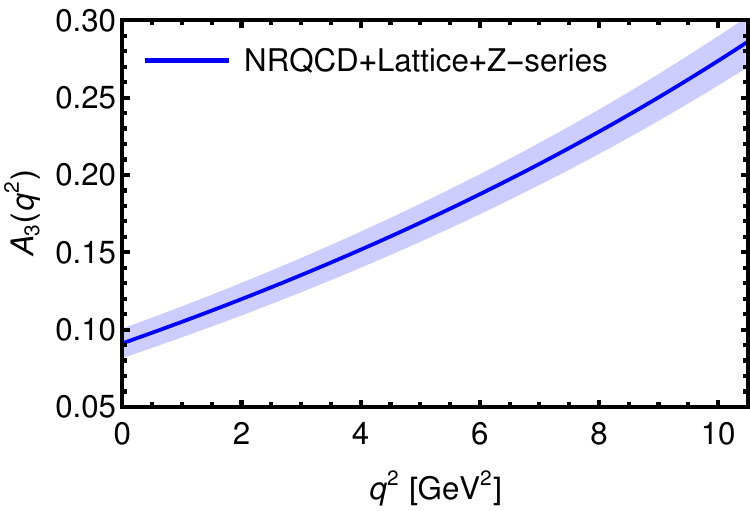}
	\caption{The same as Fig.~\ref{fig:V1q2}, but for $A_3(q^2)$.  }
	\label{fig:A3q2}
\end{figure}

\begin{figure}[!htbp]
	\centering
	\includegraphics[width=0.45\textwidth]{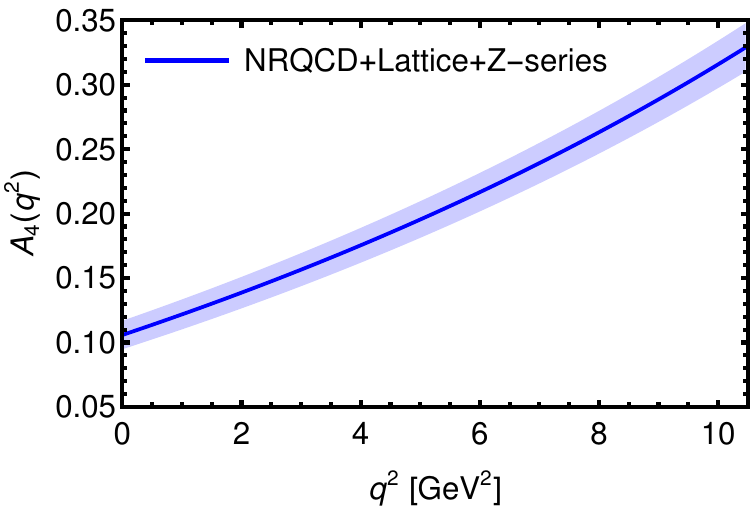}
	\caption{The same as Fig.~\ref{fig:V1q2}, but for $A_4(q^2)$.  }
	\label{fig:A4q2}
\end{figure}

\begin{figure}[!htbp]
	\centering
	\includegraphics[width=0.45\textwidth]{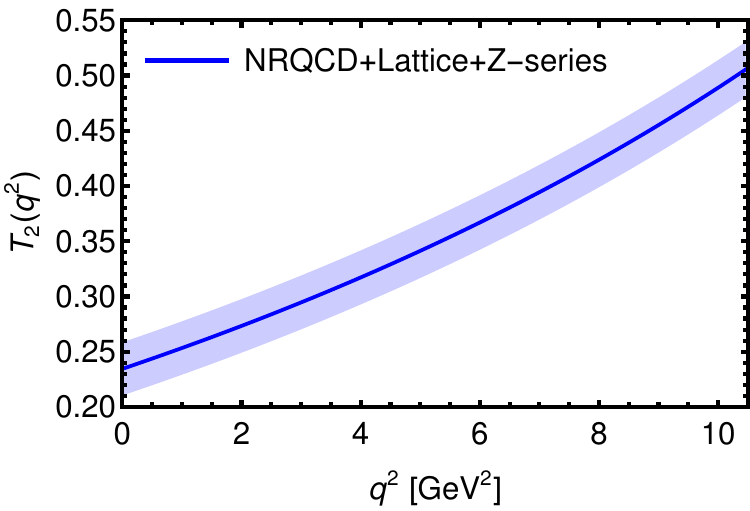}
	\caption{The same as Fig.~\ref{fig:V1q2}, but for $T_2(q^2)$.  }
	\label{fig:T2q2}
\end{figure}

\begin{figure}[!htbp]
	\centering
	\includegraphics[width=0.45\textwidth]{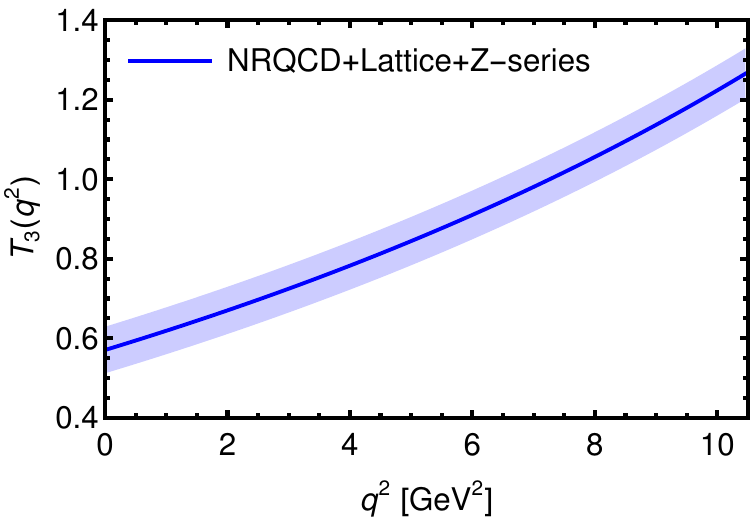}
	\caption{The same as Fig.~\ref{fig:V1q2}, but for $T_3(q^2)$.  }
	\label{fig:T3q2}
\end{figure}

\begin{figure}[!htbp]
	\centering
	\includegraphics[width=0.45\textwidth]{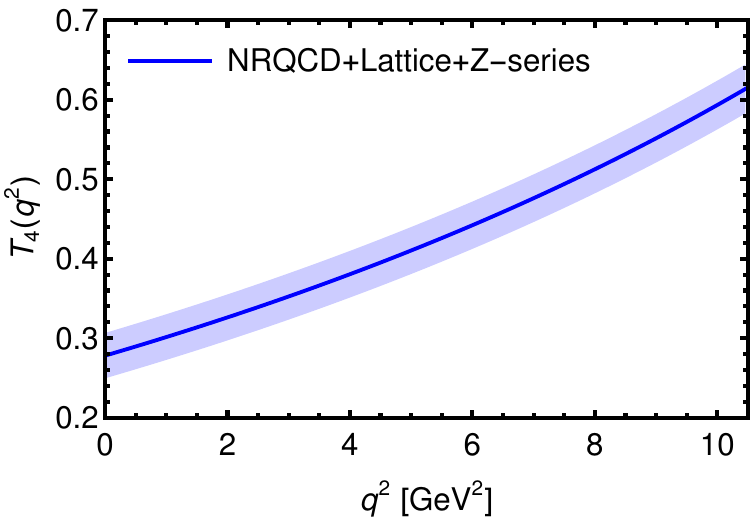}
	\caption{The same as Fig.~\ref{fig:V1q2}, but for $T_4(q^2)$.  }
	\label{fig:T4q2}
\end{figure}

\begin{figure}[!htbp]
	\centering
	\includegraphics[width=0.45\textwidth]{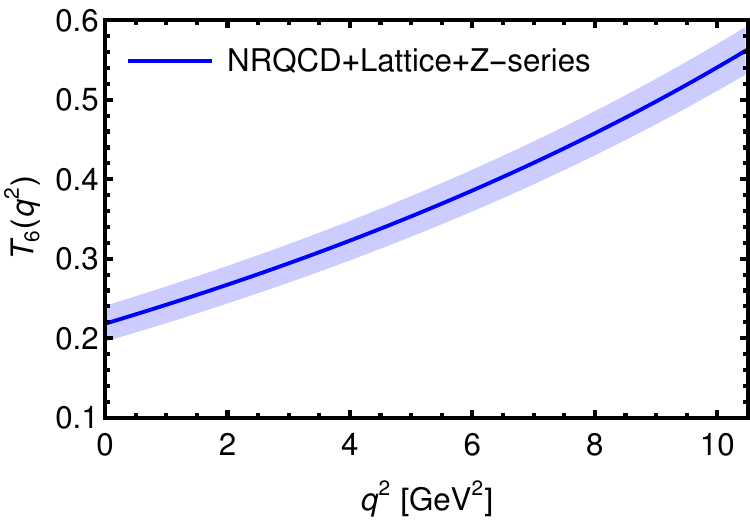}
	\caption{The same as Fig.~\ref{fig:V1q2}, but for $T_6(q^2)$.  }
	\label{fig:T6q2}
\end{figure}

\begin{figure}[!htbp]
	\centering
	\includegraphics[width=0.45\textwidth]{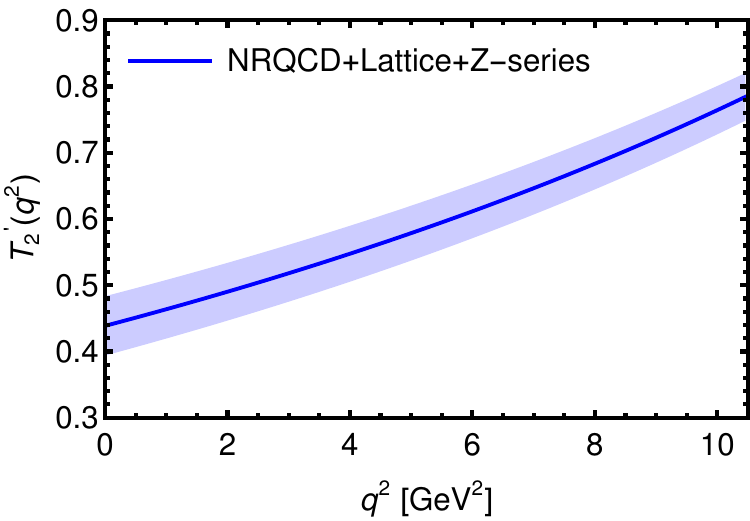}
	\caption{The same as Fig.~\ref{fig:V1q2}, but for $T'_2(q^2)$.  }
	\label{fig:AT2q2}
\end{figure}

\begin{figure}[!htbp]
	\centering
	\includegraphics[width=0.45\textwidth]{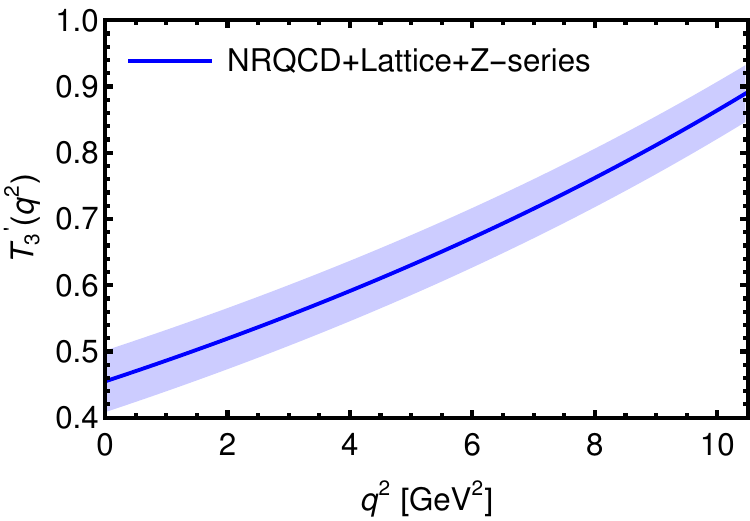}
	\caption{The same as Fig.~\ref{fig:V1q2}, but for $T'_3(q^2)$.  }
	\label{fig:AT3q2}
\end{figure}


\begin{figure}[!htbp]
	\centering
	\includegraphics[width=0.45\textwidth]{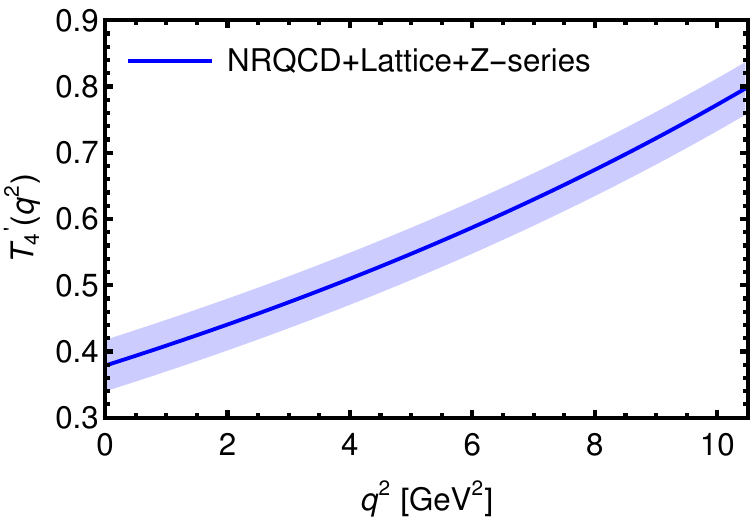}
	\caption{The same as Fig.~\ref{fig:V1q2}, but for $T'_4(q^2)$.  }
	\label{fig:AT4q2}
\end{figure}

\clearpage

To further predict the form factors at the large $q^2$ region,  we use the Z-series method~\cite{Boyd:1997kz,Bourrely:2008za,Hu:2019qcn,Leljak:2019eyw,Harrison:2020gvo,Bharucha:2010im,Hill:2006ub,Biswas:2023bqz} for the extrapolation. 
In this  method, 
the form factor is parameterized using a truncated series in powers of $z$:
\begin{align}\label{zseries}
F_{i}(q^2) =\frac{1}{1-\frac{q^2}{ m_{R}^{2}}} \sum_{n=0}^{N} \alpha_{i,n}\, z^{n}\left(q^2\right),
\end{align}
with
\begin{align}
z(q^2) &=\frac{\sqrt{t_{+}-q^2}-\sqrt{t_{+}-t_{0}}}{\sqrt{t_{+}-q^2}+\sqrt{t_{+}-t_{0}}},\\
t_{0} &=t_{+}\left(1-\sqrt{1-\frac{t_{-}}{t_{+}}}\right), \\
t_{\pm} &=\left(m_{B_{c}^*} \pm m_{J/\psi}\right)^{2},
\end{align}
where $m_R$, $m_{B_{c}^*}$ and $m_{J/\psi}$ represent the masses of the low-lying bottom-charm resonance~\cite{Boyd:1997kz,Hu:2019qcn,Leljak:2019eyw}, the $B_c^*$ meson and the $J/\psi$ meson, respectively.
We have adopted $N=1$ for the truncation, as $z(q^2)\sim 0.02$  in $B_c^*\to J/\psi$ ~\cite{Wang:2018duy,Bharucha:2010im,Leljak:2019eyw},
and the undetermined parameters $\alpha_{i,0}$ and $\alpha_{i,1}$ can be fitted by inputting the NRQCD + Lattice results in the relatively small $q^2$ region.
With the help of Eq.~\eqref{zseries}, we finally present the NRQCD + Lattice + Z-series predictions for the $B_c^*\to J/\psi$ form factors over the full physical $q^2$ range $0\leq q^2\leq (m_{B_c^*}-m_{J/\psi})^2$ in Figs.~\ref{fig:V1q2}--\ref{fig:AT4q2}, 
where the results for the (axial-)vector form factors $V_{1,2,3,4,5,6}(q^2)$ and $A_{1,2,3,4}(q^2)$ are roughly consistent with the corresponding results by  LFQM in the literature~\cite{Chang:2019obq,Chang:2020xvu}.

\section{Summary}\label{SUMMARY}

In this paper, we calculate the NLO perturbative QCD corrections to the $B_c^*\to J/\psi$ (axial-)vector and (axial-)tensor form factors  within the NRQCD factorization framework. We obtain the complete analytical results for the NLO form factors and provide their asymptotic expressions in the hierarchical heavy quark limit. 

At the maximum recoil point ($q^2=0$), we  investigate the  dependence of the LO and NLO form factors on the renormalization scale, finding that the NLO corrections reduce the scale dependence. 
Further studies reveal that the NLO corrections to the form factors are both significant and convergent in the relatively small squared transfer momentum  ($q^2$) region. 
By employing the heavy quark spin symmetry between  $B_c^*$ and $B_c$ wave functions at the origin, and combining the NRQCD results with the lattice QCD data for the $B_c\to J/\psi$ form factors, we provide the NRQCD + Lattice theoretical predictions for the $B_c^*\to J/\psi$ form factors.  Furthermore, by using the Z-series fitting method, we extend the theoretical predictions for the $B_c^*\to J/\psi$ form factors over the full physical  $q^2$ range. 

These results and predictions for the form factors can be used to calculate the decay widths and branching ratios of the semi-leptonic decays $B_c^*\to J/\psi+l\nu_l$~\cite{Chang:2020xvu,Wang:2016dkd,Wang:2018ryc,Wang:2024cyi,R:2019uyb,Sheng:2021iss}, thereby providing useful reference information for the experimental discovery of $B_c^*$. In addition, the NLO calculations for the (axial-)tensor form factors could offer valuable insights into potential new physics beyond the Standard Model.

\section*{Acknowledgements}

We thank Junfeng Sun, Yadong Yang, Jian-Ping Ma, Su-Ping Jin and Hua-Yu Jiang for many helpful discussions.
The work is supported by the National Natural Science Foundation of China (Grant No. 12275067),
Science and Technology R$\&$D Program Joint Fund Project of Henan Province  (Grant No. 225200810030),
Science and Technology Innovation Leading Talent Support Program of Henan Province,
and National Key R$\&$D Program of China (Grant No. 2023YFA1606000).
The work is partially supported by NSFC under grant No. 12322503 and 12335003.

\appendix

\section{Asymptotic NLO results of $B_c^*\to J/\psi$ (axial-)vector and (axial-)tensor form factors}\label{appendix123}
In this appendix, we present the asymptotic results for the ratios of NLO to LO form factors in the hierarchical heavy quark limit, obtained by taking the leading-order terms in the small $m_c$ expansion of the complete results. 

\begin{widetext}

In the hierarchical heavy quark limit, the asymptotic expressions read:
\begin{align}
\frac{V_1^\text{NLO}}{V_1^\text{LO}}=\,&1+\frac{\alpha_s}{4 \pi } \bigg\{\left(\frac{11 C_A}{3}-\frac{2}{3} {n_f}\right) \ln \frac{2 s {y}^2}{x}-\frac{10}{9} {n_f}+ \left(\frac{2
	\ln s}{3}-\frac{2 \ln x}{3}+\frac{10}{9}+\frac{2 \ln 2}{3}\right){n_b}
\nonumber\\&
-{C_A}\bigg[\frac{\ln^2 x}{2}+\left(\ln s+2 \ln 2+\frac{3}{2}\right) \ln x+\frac{1}{2} \ln^2 s
+\left(\frac{3}{2}+2 \ln 2\right) \ln s
+2 \ln^2 2+\frac{3 \ln 2}{2}
\nonumber\\&
-\frac{1}{9}
\left(67-3 \pi ^2\right)\bigg]
+C_F \bigg[2 \text{Li}_2(1-s)
+\ln^2 x+(2 \ln s+10 \ln 2-5) \ln x
+2 \ln^2 s
\nonumber\\&
+(10 \ln 2-2) \ln s
+7 \ln^2 2+9 \ln 2
+\frac{1}{3} \left(\pi ^2-51\right)\bigg]
\bigg\},
\end{align}

\begin{align}
\frac{V_3^\text{NLO}}{V_3^\text{LO}}=\,&\frac{V_1^\text{NLO}}{V_1^\text{LO}}+
\frac{\alpha_s }{4
	\pi }\bigg\{C_A \bigg[2 s \text{Li}_2(1-2 s)-s \text{Li}_2(1-s)+\frac{1}{2} s \ln^2 s
+\left(\frac{1}{2 s-1}+2 s
	\ln 2\right) \ln s		
	\nonumber\\&		
	+s \ln^2 2	
	+\left(\frac{1}{2 s-1}-1\right) \ln 2
	+\frac{\pi ^2 s}{6}\bigg]
	-C_F \bigg[4 s \text{Li}_2(1-2 s)-2 s \text{Li}_2(1-s)		
+s \ln^2 s	
	\nonumber\\&
-\left(\frac{2 s^2-1}{2 (1-2
		s)^2 (s-1)}-4 s \ln 2\right) \ln s
	+2 s \ln^2 2-\frac{(s (8 s-9)+3) \ln 2}{(1-2 s)^2}
	+\frac{\pi ^2 s}{3}-\frac{1}{2-4 s}\bigg]\bigg\},
\end{align}

\begin{align}
\frac{V_4^\text{NLO}}{V_4^\text{LO}}=\,&\frac{V_1^\text{NLO}}{V_1^\text{LO}}+
\frac{\alpha_s}{4 \pi } \bigg\{C_A \bigg[\left(4 s-\frac{3}{2}\right) \text{Li}_2(1-2 s)+(1-2 s) \text{Li}_2(1-s)+\frac{\ln^2 x }{4}
+\left(\frac{\ln s}{2}+\frac{5 \ln 2}{2}\right)
	\ln x
\nonumber\\&	
	+s \ln ^2 s  +\left(\frac{1}{4 s-2}+(4 s+1) \ln 2 \right) \ln s +\left(2 s+\frac{3}{2}\right) \ln ^2 2+\frac{(3-4 s) \ln
		2}{4 s-2}+\frac{1}{6} \pi ^2 (2 s-1)\bigg]
\nonumber\\&	
	+C_F \bigg[(3-8 s) \text{Li}_2(1-2 s)+(4 s-2) \text{Li}_2(1-s)-\frac{\ln ^2 x}{4} -\left(\frac{\ln s }{2}+\frac{5}{4}+2 \ln 2 \right) \ln x
		\nonumber\\&
	+\left(\frac{1}{4}-2 s\right) \ln
	^2 s 		
		+\left(\frac{-10 s^2+17 s-5}{8 s^2-12 s+4}+(1-8 s) \ln
	2\right) \ln s -\left(4 s+\frac{3}{2}\right) \ln^2 2 	
	+\left(\frac{1}{2-4 s}+6\right) \ln 2
		\nonumber\\&
	+\frac{1}{12} \left(\pi ^2 (3-8 s)+15\right)\bigg]\bigg\},
\end{align}

\begin{align}
\frac{V_5^\text{NLO}}{V_1^\text{LO}}=\,&\frac{{\alpha_s}}{4 \pi } \bigg\{{C_A} \bigg[4 s^2 \text{Li}_2(1-2 s)-2 s^2 \text{Li}_2(1-s)+s^2 \ln ^2 s+\left(4 s^2 \ln
	2+\frac{1}{2 s-1}+1\right) \ln s
			\nonumber\\&
		+2 s^2 \ln ^2 2
		+\left(-2 s+\frac{1}{2 s-1}+1\right) \ln 2+\frac{\pi ^2 s^2}{3}\bigg]+{C_F} \bigg[-8 s^2 \text{Li}_2(1-2 s)+4 s^2 \text{Li}_2(1-s)-2 s^2
	\ln ^2 s
		\nonumber\\&
	+\left(-8 s^2 \ln 2+\frac{1}{s-1}+\frac{1}{2 (1-2 s)^2}+\frac{1}{2}\right) \ln s-4 s^2 \ln ^2 2+\frac{1}{2} \left(8 s+\frac{1}{(1-2 s)^2}-1\right) \ln 2
		\nonumber\\&
	+\frac{\left(2 \pi ^2 (1-2 s) s-3\right) s}{6
		s-3}\bigg]\bigg\},
\end{align}

\begin{align}
\frac{V_6^\text{NLO}}{V_1^\text{LO}}=\,&
\frac{{\alpha_s}}{4 \pi } \bigg\{{C_A} \bigg[4 s \text{Li}_2(1-2 s)-2 s \text{Li}_2(1-s)+s \ln ^2 s+\left(\frac{2}{2 s-1}+4 s \ln
2\right) \ln s
	+2 s \ln ^2 2
+\left(\frac{2}{2 s-1}-2\right) \ln 2
\nonumber\\&
+\frac{\pi ^2 s}{3}\bigg]-{C_F} \bigg[8 s \text{Li}_2(1-2 s)-4 s \text{Li}_2(1-s)+2 s \ln ^2 s	
-\left(\frac{2 s (9 s-8)+3}{(1-2 s)^2 (s-1)}-8 s \ln 2\right) \ln s
\nonumber\\&
+4 s \ln ^2 2
-\frac{(2 s (8 s-9)+6) \ln 2}{(1-2
		s)^2}
	+\frac{2 \pi ^2 s}{3}-\frac{1}{1-2
		s}+4\bigg]\bigg\},
\end{align}

\begin{align}
\frac{T_3^\text{NLO}}{T_3^\text{LO}}=\,&\frac{V_1^\text{NLO}}{V_1^\text{LO}}+
\frac{{\alpha_s}}{4 \pi } \bigg\{{C_A} \bigg[\frac{(4-6 s) \text{Li}_2(1-2 s)}{4 s-1}+\left(\frac{3}{1-4 s}+1\right) \text{Li}_2(1-s)+\frac{\ln ^2 x}{2-8
		s}
		\nonumber\\&
	+\left(\frac{3}{2-8 s}+\frac{\ln s}{1-4 s}+\frac{6 \ln 2}{1-4 s}\right) \ln x+\frac{(2 s+1) \ln ^2 s}{2-8 s}+\left(\frac{3}{2-8 s}+\frac{(6 s+2) \ln 2}{1-4 s}\right) \ln s
		\nonumber\\&
	+\frac{(3
		s+2) \ln ^2 2}{1-4 s}+\frac{3 \ln 2}{2-8 s}-\frac{\pi ^2 (s-1)+6}{12 s-3}\bigg]+{C_F} \bigg[\left(\frac{3}{1-4
		s}+3\right) \text{Li}_2(1-2 s)+\left(\frac{4}{4 s-1}-2\right) \text{Li}_2(1-s)
		\nonumber\\&
	+\frac{2 \ln ^2 x}{4 s-1}+\left(\frac{1}{4 s-1}+\frac{4 \ln s}{4 s-1}+\frac{12
		\ln 2}{4 s-1}\right) \ln x+\frac{2 (s+1) \ln ^2 s}{4 s-1}+\left(\frac{5}{4
		s-1}+\left(\frac{9}{4 s-1}+3\right) \ln 2\right) \ln s
		\nonumber\\&
	+\frac{(6 s+5) \ln ^2 2}{4 s-1}+\frac{8 \ln 2}{4 s-1}+\frac{2 \pi ^2 (s+1)}{12 s-3}\bigg]\bigg\},
\end{align}

\begin{align}
\frac{T_4^\text{NLO}}{T_4^\text{LO}}=\,&\frac{T_3^\text{NLO}}{T_3^\text{LO}}+
\frac{{\alpha_s}}{4 \pi } \bigg\{{C_A} \bigg[\frac{\left(-8 s^2+4 s+2\right) \text{Li}_2(1-2 s)}{1-4 s}+\frac{\left(-4 s^2+2 s+2\right) \text{Li}_2(1-s)}{4 s-1}+\frac{s \ln ^2 x}{4 s-1}
\nonumber\\&
+\left(\frac{2 (5 s+1) \ln 2}{4 s-1}+\frac{2 s \ln s}{4 s-1}\right) \ln x+\frac{2 s^2
	\ln ^2 s}{4 s-1}
+2 \left(s+\frac{1}{4 s-1}+1\right) \ln 2 \ln s
+\left(\frac{8 s}{4 s-1}+s\right) \ln ^2 2
\nonumber\\&
+\frac{\pi ^2 (2 (s-1) s-1)}{12
		s-3}\bigg]+{C_F} \bigg[\left(-4 s+\frac{3}{4 s-1}+1\right) \text{Li}_2(1-2 s)+\frac{\left(-8 s^2+4 s+2\right)
		\text{Li}_2(1-s)}{1-4 s}
	+\frac{s \ln ^2 x}{1-4 s}
		\nonumber\\&
	+\left(\frac{5 s}{1-4 s}+\frac{2 s \ln s}{1-4 s}+\frac{8 s \ln 2}{1-4 s}\right) \ln x	
	-s \ln ^2 s+\left(\frac{s-10 s^2}{8 s^2-6 s+1}+\left(-4 s+\frac{1}{4 s-1}-1\right) \ln
	2\right) \ln s
		\nonumber\\&
+\frac{(1-8 s (s+1)) \ln ^2 2}{4 s-1}+\left(\frac{13}{4 s-1}+\frac{2}{1-2 s}+4\right) \ln 2	+\frac{15 s+\pi ^2
	((3-4 s) s+1)}{12 s-3}\bigg]\bigg\},
\end{align}

\begin{align}
\frac{T_6^\text{NLO}}{T_6^\text{LO}}=\,&\frac{V_1^\text{NLO}}{V_1^\text{LO}}+
\frac{{\alpha_s}}{4 \pi } \bigg\{{C_A} \bigg[(6 s-4) \text{Li}_2(1-2 s)+(4-4 s) \text{Li}_2(1-s)+\frac{\ln ^2 x}{2}+\left(\ln s+\frac{3}{2}+6 \ln 2\right) \ln x
	\nonumber\\&
+\left(s+\frac{1}{2}\right) \ln ^2 s+\left((6 s+2) \ln 2+\frac{3}{2}\right) \ln s+(3 s+2) \ln ^2 2+\frac{3 \ln
	2}{2}+\frac{1}{3} \pi ^2
(s-1)+2\bigg]
	\nonumber\\&
+{C_F} \bigg[(6-12 s) \text{Li}_2(1-2 s)+(8 s-6) \text{Li}_2(1-s)-2 \ln ^2 x-(4 \ln s+1+12 \ln 2) \ln x
-2 (s+1) \ln^2 s
	\nonumber\\&
+\left(\frac{8}{s-1}-6 (2 s+1) \ln 2+3\right) \ln s-(6 s+5) \ln ^2 2-8 \ln 2-\frac{2}{3} \pi ^2 (s+1)\bigg]\bigg\},
\end{align}

\begin{align}
\frac{V_1^\text{NLO}}{V_1^\text{LO}}=\frac{V_2^\text{NLO}}{V_2^\text{LO}}+\frac{\alpha_s C_F}{2\pi(s-1)}\left({1}-\frac{\ln s}{s-1}\right)=\frac{A_1^\text{NLO}}{A_1^\text{LO}}-\frac{\alpha_s C_F}{2\pi}\left(3-\frac{\ln s}{s-1}\right)=\frac{A_2^\text{NLO}}{A_2^\text{LO}}-\frac{\alpha_s C_F}{2\pi(s-1)}\left(3s-2-\frac{s\ln s}{s-1}\right),
\end{align}

\begin{align}
\frac{A_3^\text{NLO}}{V_1^\text{LO}}=\frac{A_4^\text{NLO}}{V_1^\text{LO}}-\frac{V_5^\text{NLO}}{V_1^\text{LO}}=\frac{\alpha_s C_F s}{2\pi(s-1)}\left({1}-\frac{\ln s}{s-1}\right),
\end{align}

\begin{align}
\frac{T_2^\text{NLO}}{T_2^\text{LO}}=\frac{{T'_2}^\text{NLO}}{{T'_2}^\text{LO}}=\frac{{T'_3}^\text{NLO}}{{T'_3}^\text{LO}}=\frac{T_6^\text{NLO}}{2T_6^\text{LO}}+\frac{{T'_4}^\text{NLO}}{2{T'_4}^\text{LO}}=\frac{V_1^\text{NLO}}{V_1^\text{LO}}+\frac{\alpha_s C_F \ln s}{2\pi(s-1)},
\end{align}
where $n_f=n_b+n_c+n_l$.

And at the maximum recoil point ($q^2=0$ or $s=1$), the above expressions reduce to:
\begin{align}
\frac{V_1^\text{NLO}}{V_1^\text{LO}}=\,&1+\frac{\alpha_s}{4 \pi } \bigg\{ \left(\frac{11 {C_A}}{3}-\frac{2}{3} {n_f}\right)\ln \frac{2 {y}^2}{x}
-\frac{10}{9} {n_f}+
\left(-\frac{2 \ln x}{3}+\frac{10}{9}+\frac{2 \ln 2}{3}\right){n_b}
\nonumber\\&
-{C_A}
	\bigg[\frac{1}{2} \ln ^2 x+\left(\frac{3}{2}+2 \ln 2\right) \ln x+2 \ln ^2 2+\frac{3 \ln 2}{2}-\frac{1}{9} \left(67-3 \pi ^2\right)\bigg]
\nonumber\\&	
	+{C_F}
	\bigg[\ln ^2 x+(10 \ln 2-5) \ln x+7 \ln ^2 2+9 \ln 2+\frac{1}{3} \left(\pi ^2-51\right)\bigg]\bigg\},
\end{align}

\begin{align}
\frac{V_4^\text{NLO}}{V_4^\text{LO}}=\,&\frac{V_1^\text{NLO}}{V_1^\text{LO}}+\frac{{\alpha_s}}{4 \pi } \bigg\{{C_A} \bigg[\frac{\ln ^2 x}{4}+\frac{5}{2} \ln 2 \ln x+\frac{7 \ln ^2 2}{2}-\frac{\ln 2}{2}-\frac{\pi ^2}{24}\bigg]
\nonumber\\&
-{C_F}
	\bigg[\frac{1}{4} \ln ^2 x+\left(\frac{5}{4}+2 \ln 2\right) \ln x+\frac{11 \ln ^2 2}{2}-\frac{11 \ln 2}{2}-\frac{7}{4}\bigg]\bigg\},
\end{align}

\begin{align}
\frac{T_3^\text{NLO}}{T_3^\text{LO}}=\,&\frac{V_1^\text{NLO}}{V_1^\text{LO}}-
\frac{{\alpha_s}}{4 \pi } \bigg\{{C_A} \bigg[\frac{1}{6} \ln ^2 x+\left(\frac{1}{2}+2 \ln 2\right) \ln x+\frac{5 \ln
		^2 2}{3}+\frac{\ln 2}{2}-\frac{1}{18} \left(\pi ^2-12\right)\bigg]
	\nonumber\\&
	-{C_F} \bigg[\frac{2 \ln ^2 x}{3}+\left(\frac{1}{3}+4 \ln 2\right) \ln x+\frac{11 \ln
		^2 2}{3}+\frac{8 \ln 2}{3}+\frac{5 \pi ^2}{18}\bigg]\bigg\},
\end{align}

\begin{align}
\frac{T_4^\text{NLO}}{T_4^\text{LO}}=\,&\frac{T_3^\text{NLO}}{T_3^\text{LO}}+
\frac{{\alpha_s}}{4 \pi } \bigg\{{C_A} \bigg[\frac{\ln ^2 x}{3}+4 \ln 2 \ln x+\frac{11 \ln ^2 2}{3}-\frac{\pi ^2}{6}\bigg]
\nonumber\\&
-{C_F} \bigg[\frac{1}{3} \ln
	^2 x+\left(\frac{5}{3}+\frac{8 \ln 2}{3}\right) \ln x+5 \ln ^2 2-\frac{19 \ln 2}{3}-\frac{1}{6} \left(10+\pi ^2\right)\bigg]\bigg\},
\end{align}

\begin{align}
\frac{V_1^\text{NLO}}{V_1^\text{LO}}=\frac{V_2^\text{NLO}}{V_2^\text{LO}}+\frac{A_3^\text{NLO}}{V_1^\text{LO}}=\frac{V_2^\text{NLO}}{V_2^\text{LO}}+\frac{\alpha_s C_F}{4\pi}=\frac{A_1^\text{NLO}}{A_1^\text{LO}}-\frac{\alpha_s C_F}{\pi}=\frac{A_2^\text{NLO}}{A_2^\text{LO}}-\frac{5\alpha_s C_F}{4\pi},
\end{align}

\begin{align}
\frac{V_5^\text{NLO}}{V_1^\text{LO}}=\frac{V_6^\text{NLO}}{V_1^\text{LO}}=\frac{2V_3^\text{NLO}}{V_3^\text{LO}}-\frac{2V_1^\text{NLO}}{ V_1^\text{LO}}=\frac{A_4^\text{NLO}}{V_1^\text{LO}}-\frac{A_3^\text{NLO}}{V_1^\text{LO}}=\frac{{\alpha_s}\ln 2}{2 \pi } [{C_A} \ln 2+2 C_F  (1-\ln 2)],
\end{align}

\begin{align}
\frac{T_2^\text{NLO}}{T_2^\text{LO}}=\frac{3 T_3^\text{NLO}}{4 T_3^\text{LO}}+\frac{T_6^\text{NLO}}{4 T_6^\text{LO}}=\frac{{T'_2}^\text{NLO}}{{T'_2}^\text{LO}}=\frac{{T'_3}^\text{NLO}}{{T'_3}^\text{LO}}=\frac{T_6^\text{NLO}}{2 T_6^\text{LO}}+\frac{{T'_4}^\text{NLO}}{2{T'_4}^\text{LO}}=\frac{V_1^\text{NLO}}{V_1^\text{LO}}+\frac{{\alpha_s}C_F}{2 \pi }.
\end{align}

\end{widetext}

\end{document}